\documentstyle[prd,aps,preprint,titlepage,psfig,amsfonts,floats,cite,array,dcolumn]{revtex}
\draft
\tighten

\newcommand{\Pomeron }{{\mathbb P}}

\begin{document}
\preprint{
        \parbox{2.0in}{%
           \noindent
           hep-ph/9805268 \\
           CTEQ-701 (rev)\\
           PSU/TH/177 (rev)\\
        }
}

\title{Diffractive Production of Jets and Weak Bosons
       and Tests of Hard-Scattering Factorization
}
\author{
   Lyndon Alvero$^{a}$, John C. Collins$^{a}$, Juan Terron$^{b}$, and
   J. J. Whitmore$^{a}$
}
\address{
   $a)\ ${\it Physics Department, Pennsylvania State University\\
              104 Davey Lab., University Park, PA 16802-6300,
              U.S.A.}
\\
   $b)\ $ {\it Universidad Aut\'onoma de Madrid,
          Departamenta de F\'\i sica Te\'orica,
          Madrid, Spain}
}

\date{30 December 1998}

\maketitle

\begin{abstract}
    We extract diffractive parton densities from data on
    diffractive deep inelastic scattering (DIS) and on
    diffractive photoproduction of jets. We explore the results
    of several ans\"atze for the functional form of the parton
    densities. Then we use the fitted parton densities to predict
    the diffractive production of jets and of $W$'s and $Z$'s in
    $p\bar p$ collisions at the Tevatron. To fit the
    photoproduction data requires a large gluon density in the
    Pomeron. The predictions for the Tevatron cross sections are
    substantially higher than data; this signals a breakdown of
    hard-scattering factorization
    in diffractive hadron-hadron collisions.
\end{abstract}

\pacs{13.60.Hb 13.60.-r  13.85.Qk  13.87.-a}



\section{Introduction}
\label{sec:intro}

In view of counterexamples \cite{nonfact,nonfact.preQCD} to the
conjecture of factorization \cite{ISorig} of hard processes in
diffractive scattering, it is important to test \cite{CTEQpom}
factorization experimentally.  In this paper, we present some results
to this end.  Specifically, we present fits\footnote{
  The fits presented in this paper represent a complete updating of
  our fits in an earlier preprint \cite{Version1}.
}
to data from the ZEUS and
H1 collaborations on diffractive deep inelastic scattering (DIS)
\cite{ZEUS.DIS,ZEUS.LPS,H1.DIS} and on diffractive photoproduction of
jets \cite{ZEUS.photo}.  Then we use these fits to predict cross
sections in hard diffractive processes in $p\bar p$ collisions, with
the assumption of factorization; we find that the predictions fail badly.

We recall that diffractive events are characterized by a large rapidity
gap, a region in rapidity where no particles are produced. We are
concerned with the case where there is a hard scattering and where the
gap occurs between the hard scattering and one of the beam
remnants. Such hard diffractive events are observed in
deep inelastic scattering (DIS) experiments
\cite{firstdiff}
and are found to have a large rate: around 10\% of the inclusive
cross section.
Diffractive jet production in $p\bar p$ collisions was earlier
reported by the UA8 collaboration \cite{UA8}, but under somewhat
different kinematic conditions (larger $|t|$)\footnote{
    By $t$ we mean the invariant momentum-transfer-squared from the
    diffracted hadron.
}.
There was also a
report of diffractive bottom production \cite{Eggert}.
Now, more diffractive data are being gathered from
a variety of lepto-hadronic
\cite{ZEUS.DIS,ZEUS.LPS,H1.DIS,ZEUS.photo} and hadronic processes
\cite{CDF.wgap,CDF.jgap,CDF.Pot,D0.ANL,D0.1,D0.2}, but with
substantially smaller fractions in the case of the diffractive
production of jets and weak vector bosons in $p\bar p$
interactions
than in DIS.

Factorization for diffractive hard scattering is equivalent to the
hard-scattering aspects of the Ingelman and Schlein model
\cite{ISorig}, where diffractive scattering is attributed to the
exchange of a Pomeron --- a colorless object with vacuum quantum
numbers. Ingelman and Schlein treat the Pomeron like a real particle,
and so they consider that a diffractive electron-proton collision is
due to an electron-Pomeron collision and that a diffractive
proton-proton collision is due to a proton-Pomeron collision.
Therefore they propose that diffractive hard cross sections are
obtained as a product of a hard-scattering coefficient (or Wilson
coefficient), a known Pomeron-proton coupling, and parton densities in
the Pomeron.

As was already known \cite{nonfact.preQCD} before the advent of QCD,
factorization is not expected to hold in general for diffractive hard
processes.  Furthermore, on the basis of a breakdown of the
triple-Regge theory for soft single-diffractive excitation,
Goulianos has proposed \cite{kgoul} to
renormalize the Pomeron flux in an energy-dependent way.  The
agreement between data and his calculated cross sections is evidence
that hard-scattering factorization is likely to break down in diffractive
hadron-hadron collisions.

However, one of us has recently proved factorization
\cite{proof} for those diffractive hard processes that are lepton
induced: these include diffractive DIS and diffractive {\em direct}
photoproduction of jets.  The proof fails for hadron-induced
processes.  In this formulation, the primary non-perturbative
quantities are diffractive parton densities \cite{KS,BS1,BS2} in the
proton.  Although we will use the terminology of ``parton densities in
the Pomeron'', this mainly gives a useful way to describe the kind
of parameterization we use for the diffractive parton densities,
together with an indication of the quantum numbers that we believe to
be exchanged across the rapidity gap.  There is no necessary
requirement that the object we call
the Pomeron is the same as in soft scattering\footnote{
   So Dokshitzer \cite{Dokshitzer} would probably object to our use
   of the word ``Pomeron''.
}.

In principle,
the parton densities in the Pomeron can be extracted from
diffractive DIS $(F_{2})$ measurements alone.  Since the Pomeron is
isosinglet and is its own charge conjugate, there is only a
single light quark density to measure; one does not have the
complications of separating the different flavors of quark that
one has in the case of the parton densities of
the proton.  The $Q$ dependence of the structure functions enables one
to determine the gluon density. The H1 collaboration has already
presented \cite{H1.DIS} a fit of this kind. This type of data
sufficiently determines the quark density in the Pomeron, and the
H1 fits suggest a large gluon content for the Pomeron.
However, a more
direct measurement of the gluons can be made in photoproduction,
since the leading order processes have both quark- and
gluon-induced terms.  The ZEUS collaboration has already presented
experimental evidence for a large gluon content of the Pomeron; they
performed a combined analysis of their results on the diffractive
structure function in deep inelastic scattering \cite{ZEUS.DIS} and on
diffractive jet photoproduction \cite{ZEUS.photo}.

The main result of the ZEUS work was information on the overall
normalization of the diffractive parton densities.  In this paper, our
aim is to obtain more detailed fits including the H1 data, and to use
the resulting fits to predict other cross sections.  We use data on
both DIS and photoproduction.  Recently, the ZEUS collaboration has
reported \cite{ZEUS.Jerusalem} new fits to their data that are made
independently, but in a similar fashion to ours.

For fitting the DIS data, we use full next-to-leading-order (NLO)
calculations.  The use of NLO rather than LO calculations is important
since the gluon density is larger than the quark density.  For
the photoproduction data, we use leading-order calculations in a
Monte-Carlo event generator in order to implement the experimental
cuts.  The event generator was constructed by two of us
\cite{POMPYT-C} as an extension to the POMPYT generator to allow the
use of evolved parton densities in the Pomeron.  With the resulting
diffractive parton densities we calculate hard diffractive processes
in hadron-hadron collisions, given the assumption of factorization.

In the past, Ingelman and Schlein \cite{ISorig} and
Bruni and Ingelman \cite{BI} have made similar calculations for
one of the hadron-induced processes that we consider here ($W/Z$ production).
Their results have provided a commonly used benchmark in the
phenomenology of these processes.
They provide a
choice of either ``hard'' or ``soft'' distributions of partons in the
Pomeron, according to the $\beta \to 1$ behavior.\footnote{
   Here, $\beta $ is the fraction of the Pomeron's momentum that is
   carried by the struck parton.
}
The hard distributions give larger diffractive cross sections. At
that time, there were no data to determine the distributions. We
will find that although the quark distributions preferred by the
DIS data are hard, our cross sections are substantially below
those predicted by Bruni and Ingelman.
We will present an analysis of the reasons for the lower values that
we find.

Nevertheless, our predictions for hadron-induced cross sections
are well above the measurements
\cite{CDF.wgap,CDF.jgap,CDF.Pot,D0.ANL,D0.1,D0.2}, for both $W$
production and jet production.  In the case of jet production,
the excess only occurs because of the large gluon density
that is strongly preferred by the photoproduction data.

This paper is organized as follows.  In section \ref{sec:fits},
we show our fits to diffractive deep inelastic and photoproduction
data.
In section
\ref{sec:kin}, we present some details of the formulae used to
calculate the cross sections in hadron-hadron processes, and we
discuss the kinematics and phase-space cuts that we used.  Then in
sections \ref{sec:VB.calcs} and \ref{sec:jet.calcs}, we present and
discuss the results obtained for vector boson production and jet
production, respectively.  Finally, we summarize our findings in
section \ref{sec:concl}.

Other fits to the diffractive structure functions measured by H1
have been made by Gehrmann and Stirling \cite{GS} and by Kunszt
and Stirling \cite{KS}.
Golec-Biernat and Kwieci\'nski \cite{GK} assumed a
parameterization of the parton densities in the Pomeron and found
it to be compatible with the H1 data on diffractive DIS.
Their quark densities are about 30\% smaller than ours, and
they required the momentum sum rule to be valid.
The new features of our work are a fit to a wider range of data,
including photoproduction, the lack of an assumption of the momentum
sum rule, and a calculation of the cross
sections for diffractive jet and $W$ and $Z$ production,
so as to test factorization by comparison with data from the CDF and
D0 experiments.

\section{Partons in the Pomeron}
\label{sec:fits}

We will present a series of fits of parton densities in the
Pomeron to data on diffractive DIS and diffractive
photoproduction of jets.
There are four sets of data that we use:
\begin{itemize}

\item DIS data obtained by ZEUS using the rapidity gap
    method\cite{ZEUS.DIS};

\item DIS data obtained by ZEUS using their leading proton
    spectrometer (LPS) \cite{ZEUS.LPS};

\item DIS data obtained by H1 using the rapidity gap
    method\cite{H1.DIS}; and

\item photoproduction data obtained by ZEUS using the rapidity
    gap method\cite{ZEUS.photo}.

\end{itemize}

\subsection{DIS}
\label{sec:DIS}

Diffractive structure functions are related to the differential
cross section for the process $e + p \to  e + p + X$:
\begin{equation}
   \frac {d^{4}\sigma _{\rm diff}}{d\beta  dQ^{2} dx_{\Pomeron} dt}
   = \frac {2\pi \alpha ^{2}}{\beta Q^{4}}
     \left\{
        \left[ 1 + (1-y)^{2} \right] F_{2}^{D(4)}
        - y^{2} F_{L}^{D(4)}
     \right\} ,
\end{equation}
where corrections due to $Z^{0}$ exchange and due to radiative
corrections have been ignored.  Here $x_{\Pomeron}$ is the
fractional momentum loss of the diffracted proton (in the sense
of light-cone momentum), and $t$ is the invariant momentum
transfer for the diffracted proton.
The variables $Q^{2}$ and $y$ are the usual DIS variables, and
$\beta =x_{\rm bj}/x_{\Pomeron}$, with $x_{\rm bj}$ being the usual Bjorken
scaling variable of DIS.

Except for the ZEUS LPS data, the momentum transfer $t$ is not
measured, so we make fits to the structure function integrated over
$t$, and write the structure function in the form:
\begin{eqnarray}
    F_{2}^{D(3)}(\beta ,Q^{2},x_{\Pomeron})
    &=& \int _{-1}^{0} dt \,
      F_{2}^{D(4)}(\beta ,Q^{2},x_{\Pomeron},t) .
\label{eq.dlflux}
\end{eqnarray}
(We have set the lower limit on $t$ to $-1 \, {\rm GeV}^2$ to avoid
including contributions where the putative diffracted proton results
from fragmentation of a high $p_T$ jet.
This point should not be important at small $x_\Pomeron$.  Moreover,
the integrand in Eq.~(\ref{eq.dlflux}) is steeply falling in $t$ so
that the contributions to the integral from the region $t<-1$
are quite small.)

We next use hard-scattering factorization, proved in \cite{proof}, to
write the diffractive structure function in terms of diffractive
parton densities and hard-scattering coefficients:
\begin{equation}
    F_{2}^{D(3)}(\beta ,Q^{2},x_\Pomeron)
    = \sum _{a} e_{a}^{2} \beta  f_{a}^{D(3)}(\beta, Q^2, x_\Pomeron)
      + \mbox{NLO corrections} ,
\label{DIS.factorization}
\end{equation}
an equation valid to the leading power in $Q$.  The hard-scattering
coefficients are the same as in ordinary inclusive DIS. The predictive
power of this equation comes from the DGLAP evolution equation obeyed
by the parton densities and from the universality of the parton
densities: they can be used to predict the cross sections for certain
other diffractive hard processes.  Factorization also holds for the
diffractive structure function differential in $t$.

We now assume that $x_\Pomeron$ is small.  It is therefore sensible to
use a parameterization of the $x_\Pomeron$ dependence that is
motivated by Regge theory.

If Regge factorization is valid, then the dependence on $x_\Pomeron$
is of the form given by Regge theory, and therefore can be represented
by a Pomeron flux factor, $f_{{\Pomeron}/{p}}$, that is related
to the Pomeron-proton coupling measured in proton-proton elastic
scattering.  We do not necessarily expect Regge factorization to be
valid.  Nevertheless, we will assume that a suitable parameterization
of the $x_\Pomeron$-dependence is of the Regge form, but possibly with
different parameters than in proton-proton elastic scattering.  If
this form is not suitable, then we will find that we cannot fit the
data, and a more general parameterization is needed.  This can happen
even though hard-scattering factorization remains valid in the form,
(\ref{DIS.factorization}), proved in Ref.\ \cite{proof}.

So we will write the diffractive parton densities as a Pomeron flux
factor times what are termed parton densities in the Pomeron:
\begin{eqnarray}
    f_{a}^{D(3)}(\beta ,Q^{2},x_{\Pomeron})
    &=& f_{{\Pomeron}/{p}}(x_{\Pomeron})
      f_{a/\Pomeron}(\beta ,Q^{2}).
\end{eqnarray}
Furthermore we will assume the Pomeron flux factor is of the
Donnachie-Landshoff (DL) \cite{DLflux} form:
\begin{equation}
   f_{{\Pomeron}/{p}}^{\rm DL}(x_{\Pomeron})=
   \int _{-1}^{0} dt \, \frac {9\beta _{0}^{2}}{4\pi ^{2}}
   \biggl[{4m_{p}^{2}-2.8t\over 4m_{p}^{2}-t}
      \biggl({1\over 1-t/0.7}\biggr)^{2}
   \biggr]^{2}x_{\Pomeron}^{1-2\alpha (t)},
\label{DLflux}
\end{equation}
where $m_{p}$ is the proton mass, $\beta _{0}\simeq 1.8\ {\rm
GeV}^{-1}$ is the Pomeron-quark coupling and $\alpha
(t)=\alpha_{\Pomeron}+0.25t$ is the Pomeron trajectory.  We treat
$\alpha_{\Pomeron}$ as a parameter of our fits, instead of using the
value given by Donnachie and Landshoff.  Up to logarithmic
corrections, the flux factor integrated over $t$ is
\begin{equation}
   f_{\Pomeron/p}(x_{\Pomeron})
   \simeq C x_{\Pomeron}^{1-2\alpha _{\Pomeron}},
\label{Pomeron.Power}
\end{equation}
where $C$ is a constant.

The $t$-dependence of the DL flux factor is not used in any of our
fits, so the only use we make of the $t$ dependence in Eq.\
(\ref{DLflux}) is to give a convention for a normalization factor that
is convenient for comparisons with other work.

There is in fact another Pomeron flux factor that is commonly used,
that of Ingelman and Schlein (IS) \cite{ISorig}.  This differs from
the DL flux factor primarily in its normalization.  Since the same
normalization factor appears in all our cross sections, its value
is irrelevant to our phenomenology.  Any change in the
normalization factor is completely compensated by changing the parton
densities by an inverse factor, and we obtain the parton densities
from fitting a set of data without any {\it a priori} expectations as
to their normalization.

However, the normalization does affect the question of whether
the momentum sum rule is obeyed by the parton densities in the
Pomeron.  Since it is not at present understood whether the sum
rule is a theorem, this issue will not affect us.  The momentum
sum rule is {\em not} assumed in any of our fits.

We will use Regge theory to make one further (correctable) assumption;
this is, in effect, that the Pomeron has a definite charge conjugation
parity and is an isosinglet.  This implies that the distributions of $u$,
$d$, $\bar u$ and $\bar d$ quarks are equal.  Such an assumption is
also valid in simple models where the rapidity gap is generated by
gluon exchange. One possible mechanism for violation of the equality
of the light parton densities would be the existence of an odderon,
which has opposite charge conjugation to the Pomeron.  The existence
of Pomeron-odderon interference would break the equality of the quark
and antiquark distributions.  We will ignore this possibility, since
there is no convincing phenomenological evidence to persuade us of the
odderon's existence. We also note that the issue does not concern us
in DIS and photoproduction, since the hard-scattering coefficients are
the same for quarks and antiquarks; in effect we will measure the
average of the quark and antiquark distributions.  Odderon contributions
would only matter when we make predictions for diffractive cross
sections at the Tevatron.

In the data obtained using the rapidity gap method
\cite{ZEUS.DIS,H1.DIS}, the outgoing proton is not detected.
Such data include ``double-dissociative'' contributions where the
proton is excited to a state that escapes down the beam-pipe and thus,
misses the detector.
Factorization works for such final states, but
since we will also wish to fit data where the outgoing proton is
detected, we prefer to correct the data to remove the
double-dissociative contribution.  In the case of the ZEUS rapidity
gap data \cite{ZEUS.DIS}, excited states up to about 4 GeV pass the
diffraction selection cuts,
and it is estimated that there
is a contribution of $(15\pm 10)\%$ to the measured diffractive
$F_{2}$ from double-dissociative events.  In obtaining our fits, we
have corrected the relevant ZEUS data to take this into account.
For the case of photoproduction data, we make the corresponding corrections
for double dissociation and for nondiffractive contributions as well.
No corrections have been made to the H1 diffractive $F_2$ data for which
excited states up to 1.6 GeV are included.
This point is relevant
when we compare predictions obtained using our fits to data where the
diffracted proton is detected (as in Sect.\ \ref{sec:jet.calcs}) and
also when we later compare our fits to both ZEUS and H1 data.

\subsection{Photoproduction of Jets}

Similar formulae apply to photoproduction.  For the direct
diffractive photoproduction of a jet, $\gamma + p \to {\rm jet} + X +
p$, we let $E_T$ and $\eta$ be the transverse energy and
pseudorapidity of the jet.  Then the cross section is
\begin{equation}
    \frac{d\sigma}{dx_\Pomeron  dt  dE_T  d\eta}
    = \sum_a \int d\beta
      \, f_{\Pomeron/p}(x_\Pomeron,t)
      \, f_{a/\Pomeron}(\beta)
      \frac{d\hat\sigma_{\gamma+a \to J+X}}{dE_T  d\eta} .
\end{equation}
Here, $d\hat\sigma_{\gamma+a \to J+X}$ is the hard-scattering
coefficient for the production of a jet in the collision of a photon
and a parton of type $a$.  It is the same as in inclusive
photoproduction.  The Pomeron flux factor $f_{\Pomeron/p}$ and the parton
densities in the Pomeron are the same as in the previous section.

The proof \cite{proof} of the factorization theorem indicates that
factorization is valid for the direct part but probably not for the
resolved part of diffractive photoproduction of jets.
Fortunately, most of the cross section is from the direct process.
This is known from the experimental data \cite{ZEUS.photo}, and is
also verified by our Monte-Carlo calculations.
For the kinematic configurations of the data, we find a direct
contribution that is 2 to 4.5 times larger than the resolved
contribution, except at $\eta=0.75$, where the two terms
are comparable in size.

If we use the factorization formula to calculate cross sections for
diffractive photoproduction, then presumably we should multiply the
resolved contribution to the cross section by a correction
factor\footnote{
   On the basis of experimental evidence and of Regge models
   \cite{nonfact.preQCD} for diffractive scattering, we might expect
   the correction factor to be less than unity, a suppression factor.
   However, the coherent Pomeron mechanism of Collins, Frankfurt and,
   Strikman  \cite{nonfact} would enhance the cross section.  We will
   discuss this issue further in the conclusions.
}
similar to the one needed in hadron-hadron scattering (Secs.\
\ref{sec:VB.calcs} and \ref{sec:jet.calcs}).  Given the dominance of the
direct contribution and the low precision of the current data ---
Fig.\ \ref{photo.fits} --- we feel that this is an unnecessary
refinement in the present work.

Beyond leading order, the separation between the resolved and direct
processes is not unambiguous.  Again, at the level of accuracy of the
present data, we think that this is not an important enough issue to
affect our results.

\subsection{Selection of data}
\label{sec:selection}

As far as the DIS data are concerned, we restrict our attention to the
subset of the data that is in the truly
diffractive region.  So
we now explain the criteria we use to select the data for our fits.

For the purposes of this paper, we define the diffractive component of
a cross section to be the part of the cross section corresponding to
the leading-power $x_\Pomeron$-dependence, of the form in Eq.\
(\ref{Pomeron.Power}).  With this definition of diffraction, the cross
sections reported by the ZEUS experiment
\cite{ZEUS.DIS, ZEUS.LPS, ZEUS.photo} are the diffractive
components.

We do not need to address the question of whether the
power dependence we use, with $\alpha_\Pomeron$ around 1.1 or 1.2, is
the ultimate asymptotic behavior as $x_\Pomeron \to 0$.  We also do
not require that this power be the same as in soft diffraction.  It is
sufficient that the power law represents an adequate approximation to
a measurable part of the cross section, given that the factorization
theorem \cite{proof} applies quite generally, and not just at small
$x_\Pomeron$.  This in fact implies that our restriction to
diffractive data is mainly a matter of convenience, to reduce the
number of parameters, and to be in a region of $x_\Pomeron$ common to
the four sets of data to which we make fits.

The H1 data \cite{H1.DIS} include both non-diffractive and diffractive
components, as is evidenced by the experiment's fit to their data with
two powers of $x_\Pomeron$.  To restrict our own fit to the
diffractive region, we imposed the following cuts on all the DIS data:
at $\beta = 0.175$ or 0.2, we require $x_\Pomeron < 2\times10^{-3}$,
at $\beta = 0.375$ or 0.4, we require $x_\Pomeron < 4\times10^{-3}$,
and
at $\beta = 0.65$, we require $x_\Pomeron < 1\times10^{-2}$.
We estimated these cuts by examining where the power-law associated
with the Pomeron dominates H1's fits to the $x_\Pomeron$ dependence.
The H1 data at $\beta=0.1$ and $\beta=0.04$
are eliminated from
our fits by this criterion.

Another significant
constraint
is that we must restrict our fits to
the truly deep-inelastic region.  Outside of this region, the
leading-twist QCD factorization theorem for DIS does not hold.  In
fact some of the H1 data lie very much in the resonance region.  For
example, they have points at $\beta=0.9$ and $Q^2 = 4.5 \, {\rm
GeV}^2$.  There the invariant mass of the excited hadronic system is
$m_X = Q \sqrt{1/\beta-1} = 0.7 \, {\rm GeV}$, i.e., close to the
$\rho$ resonance.  While there are perturbative QCD results that apply
in this region, they certainly do not include the usual inclusive
factorization formula, Eq.\ (\ref{DIS.factorization}).  Most of H1's
data at $\beta = 0.9$ are at low $m_X$, while the data at smaller
$\beta$ have $m_X$ greater than about 2 GeV.  Therefore we have simply
chosen to discard the $\beta = 0.9$ data when we make our fits.

With these cuts, the set of data which we fit comprises 77 points, of
which 22 are from ZEUS DIS data obtained with the rapidity gap method
\cite{ZEUS.DIS}, 3 from ZEUS DIS data from its LPS \cite{ZEUS.LPS}, 48
from H1 DIS data \cite{H1.DIS}, and 4 from the ZEUS photoproduction
data \cite{ZEUS.photo}.  These subsets of data we will call ``ZEUS
F2D3'', ``ZEUS LPS'', ``H1 F2D3'', and ``ZEUS Photo'', respectively.

The region in which we make the fits is shown in Figs.\ \ref{H1.fits},
\ref{ZEUS.fits} and \ref{photo.fits} below, which compare our fits
to the data used to make the fits.

\subsection{Fits}

Each of our fits is represented by a parameterization of the initial
distributions at $Q_{0}^{2} = 4 \, {\rm GeV}^{2}$ for the $u$, $\bar
u$, $d$, and $\bar d$ quarks and for the gluon.  The other quark
distributions are assumed to be zero at this scale.  For the DIS cross
sections, we used full NLO calculations (with full evolution and with
the number of flavors set equal to 3), while for the photoproduction
cross sections, we used a version of POMPYT that two of us have constructed
\cite{POMPYT-C}, with the same evolved parton distributions as we used
for DIS.
The factorization and renormalization scheme is $\overline{{\rm MS}}$
here and throughout this paper.
As stated above, the Pomeron flux factor is of the Donnachie-Landshoff
form Eq.\ (\ref{DLflux}), but with an adjustable parameter for
$\alpha_\Pomeron$, and  we did not assume a momentum sum rule for the
parton densities in the Pomeron, so that the choice of normalization
for the flux factor is irrelevant.  The fits were made by minimizing
$\chi^2$, with the experimental systematic errors being added in
quadrature to the statistical errors; no attempt was made to handle
point-to-point correlated errors.  The program used to perform the
evolution was that of CTEQ \cite{CTEQ}.

We tried five functional forms for the parton densities, which we
label A, B, C, D, and SG.  For each of these five forms (Eqs.\
(\ref{paramABCD}) and (\ref{paramSG}), below), we present the values
of the parameters that give the best fit, given in turn each of the
following three values of $\alpha_\Pomeron$:
\begin{itemize}
\item $\alpha_\Pomeron = 1.08$, which represents an appropriate
    value for a conventional Pomeron, as seen in soft scattering;
\item $\alpha_\Pomeron = 1.14$, which approximates the best value
    of $\alpha_\Pomeron$ associated with any of the parameterizations
    except D; and
\item $\alpha_\Pomeron = 1.19$, which gives the best fit associated
    with parameterization D.
\end{itemize}
Since it is time-consuming to generate Monte-Carlo events
for the photoproduction process and since the number of
photoproduction data is small, we first made some preliminary fits to
DIS data alone to determine suitable values for
$\alpha_\Pomeron$, as listed above.  Since the $\chi^2$ is not
strongly dependent on $\alpha_\Pomeron$, this seems to us to be
sufficient. We will comment on the numerical values later.

Four of the parameterizations, labeled A, B, C and D, use conventional
shapes for the initial distributions.  The final fit has a gluon
distribution that is peaked near $\beta =1$, as suggested by the fit
\cite{H1.DIS} obtained by the H1 collaboration; we call this our
``super-hard gluon'' fit, SG.

\begin{table}
\begin{center}
\begin{tabular}{|l!{\vrule width 2pt}c|c|c|}
\multicolumn{2}{l}{}\\
\multicolumn{3}{l}{$\alpha_\Pomeron=1.08$:} \\
\hline
      Fit  & $a_q$             & $a_g$          & $\tilde a_q$       \\
    \hline
      A    & $0.496 \pm 0.013$ &  0             &  0                 \\
      B    & $0.493 \pm 0.013$ & $9.3 \pm 2.5$  &  0                 \\
      C    & $0.501 \pm 0.022$ &  0             & $-0.008 \pm 0.031$ \\
      D    & $0.565 \pm 0.026$ & $15.4 \pm 3.1$ & $-0.113 \pm 0.031$ \\
      SG   & $0.470 \pm 0.015$ & $12.6 \pm 3.9$ &  0                 \\
\hline
\multicolumn{2}{l}{}\\
\multicolumn{3}{l}{$\alpha_\Pomeron=1.14$:} \\
\hline
      Fit  & $a_q$             & $a_g$           & $\tilde a_q$       \\
    \hline
      A    & $0.240 \pm 0.006$ &  0              &  0                 \\
      B    & $0.239 \pm 0.006$ & $4.5  \pm 0.5$  &  0                 \\
      C    & $0.249 \pm 0.011$ &  0              & $-0.031 \pm 0.029$ \\
      D    & $0.292 \pm 0.013$ & $ 9.7 \pm 1.7$  & $-0.159 \pm 0.029$ \\
      SG   & $0.225 \pm 0.008$ & $ 7.4 \pm 2.2$  &  0                 \\
\hline
\multicolumn{2}{l}{}\\
\multicolumn{3}{l}{$\alpha_\Pomeron=1.19$:} \\
\hline
      Fit  & $a_q$             & $a_g$           & $\tilde a_q$       \\
    \hline
      A    & $0.136 \pm 0.004$ & 0               &  0                 \\
      B    & $0.135 \pm 0.004$ & $2.6  \pm 0.6 $ &  0                 \\
      C    & $0.143 \pm 0.006$ &  0              & $-0.042 \pm 0.028$ \\
      D    & $0.175 \pm 0.008$ & $6.7  \pm 1.0 $ & $-0.191 \pm 0.026$ \\
      SG   & $0.126 \pm 0.005$ & $5.0  \pm 1.4 $ &  0                 \\
\hline
\end{tabular}
\bigskip
\caption{\sf Parameters of the fits for three different values of
         $\alpha_\Pomeron$.}
\label{Params}
\end{center}
\end{table}

Parameterizations A--D are all of the general form
\begin{eqnarray}
    \beta f_{q/\Pomeron}(\beta ,Q_{0}^{2})
&=&
        a_q \left[ \beta  (1-\beta )
                  + \tilde a_q \, (1-\beta )^{2}
            \right],
\nonumber \\
    \beta f_{g/\Pomeron}(\beta ,Q_{0}^{2})
&=&
        a_g \beta  (1-\beta  ) ,
\label{paramABCD}
\end{eqnarray}
with a series of constraints on the parameters.  Note that since the
Pomeron is isosinglet and self-charge-conjugate, the distributions of
the $u$, $d$, $\bar u$, and $\bar d$ quarks are all equal. Our first
parameterization A, represents a conventional hard quark
parameterization, where we set $\tilde a_q = a_g =0$.  Then in
parameterization C we allow a soft quark term, while keeping no gluon
term, so that $a_g=0$.  In parameterization B we allow an initial
gluon distribution but do not allow a soft quark term, so that $\tilde
a_q = 0$.  Finally, in parameterization D we remove all the
constraints.

The super-hard gluon parameterization, SG, is of the form
\begin{eqnarray}
    \beta f_{q/\Pomeron}(\beta ,Q_{0}^{2})
&=&
        a_q \beta  (1-\beta ) ,
\nonumber \\
    \beta f_{g/\Pomeron}(\beta ,Q_{0}^{2})
&=&
        a_g \beta^8  (1-\beta  )^{0.3} ,
\label{paramSG}
\end{eqnarray}
i.e., the quark has a hard form, and the gluon is strongly emphasized
at large $\beta$.  The exponents for the gluon distribution were
chosen somewhat arbitrarily, with no attempt being made to fit them.

\begin{table}
\begin{center}
\begin{tabular}{|l!{\vrule width 2pt}*{5}{D{.}{.}{-1}|}}
\multicolumn{2}{l}{}\\
\multicolumn{3}{l}{$\alpha_\Pomeron=1.08$:} \\
\hline
Fit &
    \multicolumn{1}{l|}{Zs F2D3} &
    \multicolumn{1}{l|}{Zs LPS} &
    \multicolumn{1}{l|}{H1 F2D3} &
    \multicolumn{1}{l|}{Zs Photo} &
    \multicolumn{1}{l|}{All Sets} \\
\hline
A   &  8.2   &  1.8   & 81.9   &  9.9   & 101.8 \\
B   &  5.9   &  2.0   & 77.7   &  2.1   &  87.8 \\
C   &  8.5   &  1.8   & 81.6   &  9.9   & 101.8 \\
D   &  9.3   &  1.8   & 65.3   &  1.2   &  77.6 \\
SG  &  6.6   &  1.9   & 80.7   &  2.1   &  91.2 \\
\hline
\multicolumn{2}{l}{}\\
\multicolumn{3}{l}{$\alpha_\Pomeron=1.14$:} \\
\hline
Fit &
    \multicolumn{1}{l|}{Zs F2D3} &
    \multicolumn{1}{l|}{Zs LPS} &
    \multicolumn{1}{l|}{H1 F2D3} &
    \multicolumn{1}{l|}{Zs Photo} &
    \multicolumn{1}{l|}{All Sets} \\
\hline
A   &  8.6   & 3.3   & 68.8   & 10.1   &  90.8 \\
B   &  5.8   & 3.4   & 65.3   &  3.3   &  77.8 \\
C   &  9.7   & 3.2   & 66.8   & 10.1   &  89.7 \\
D   & 10.8   & 2.4   & 42.1   &  1.1   &  56.3 \\
SG  &  6.2   & 3.7   & 67.7   &  1.9   &  79.6 \\
\hline
\multicolumn{2}{l}{}\\
\multicolumn{3}{l}{$\alpha_\Pomeron=1.19$:} \\
\hline
Fit &
    \multicolumn{1}{l|}{Zs F2D3} &
    \multicolumn{1}{l|}{Zs LPS} &
    \multicolumn{1}{l|}{H1 F2D3} &
    \multicolumn{1}{l|}{Zs Photo} &
    \multicolumn{1}{l|}{All Sets} \\
\hline
A   &  9.5   &  5.0   & 72.3      & 10.2   &  97.1 \\
B   &  6.3   &  4.9   & 68.3      &  4.0   &  83.6 \\
C   & 11.0   &  4.7   & 69.0      & 10.2   &  95.0 \\
D   & 12.4   &  3.0   & 34.8      &  1.0   &  51.2 \\
SG  &  6.4   &  5.8   & 70.6      &  2.0   &  84.9 \\
\hline
\end{tabular}
\end{center}
\bigskip
\caption{\sf $\chi^2$ for each of the fits.
     The data sets and the number of points are:
     ZEUS F2D3: 22 points, ZEUS F2D3 LPS: 3 points,
     H1 F2D3: 48 points, ZEUS photoproduction: 4 points.
     The total number of data points is 77.
}
\label{Chi2.Table}
\end{table}

\begin{figure}
\centerline{\psfig{file=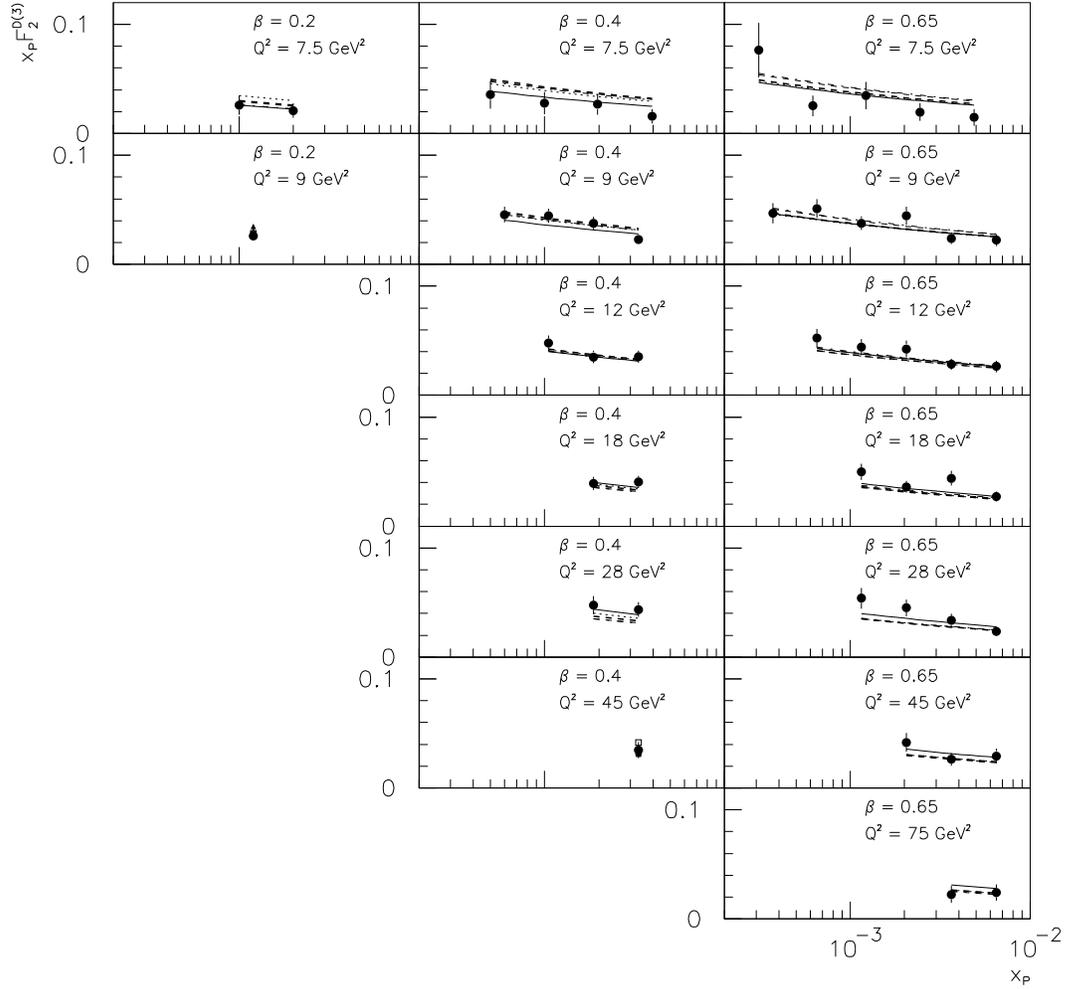,height=6in,clip=}}
\caption{\sf Comparison of the fits for $\alpha_\Pomeron = 1.14$ and the
    DIS data from H1 \protect\cite{H1.DIS} that were used in the
    fits.
    Fit A is represented by the dashed line, fit B by the dotted line,
    fit C by the dot-dashed line, fit D by the solid line, and fit SG
    by the heavy dashed line.
}
\label{H1.fits}
\end{figure}

\begin{figure}
\centerline{\psfig{file=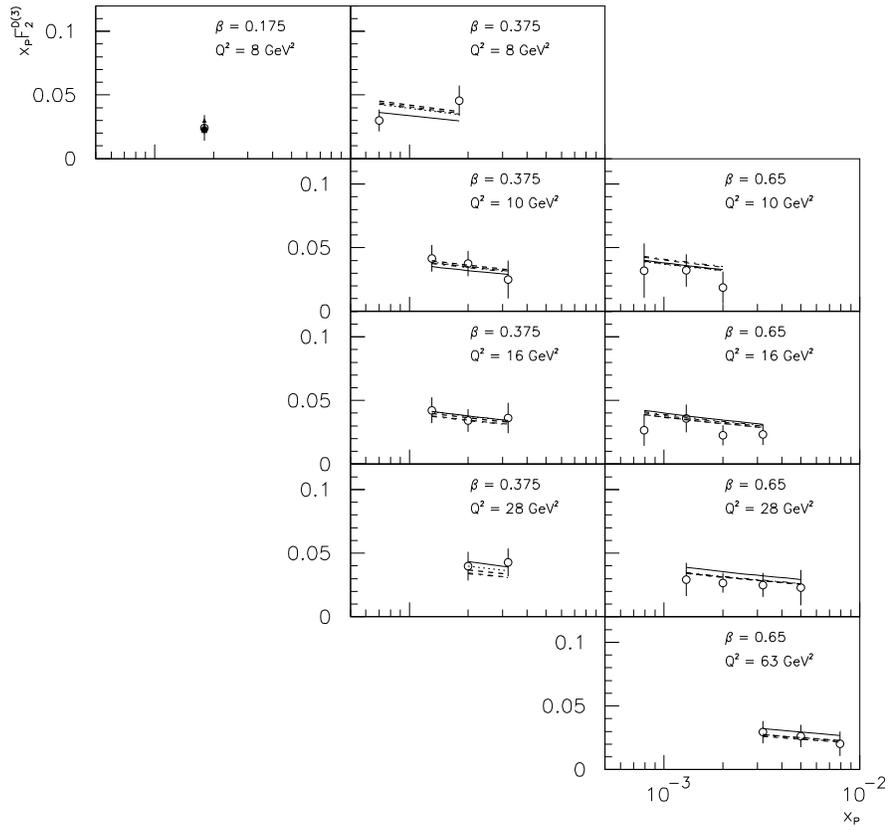,height=5in,clip=}}
\caption{\sf Comparison of the fits for $\alpha_\Pomeron = 1.14$ and the
    DIS data from ZEUS \protect\cite{ZEUS.DIS,ZEUS.LPS} that were
    used in the fits.  The LPS data we used consist of just the three
    points at $Q^2 = 8 \, {\rm GeV}^2$.
    The code for the lines is the same as in Fig.\
    \protect\ref{H1.fits}.
}
\label{ZEUS.fits}
\end{figure}

In Table \ref{Params}, we show the parameters for
each of the fits, and in Table \ref{Chi2.Table} we show the values of
$\chi^2$, both for the total set of data and for each of the four
subsets separately.
In Figs.\ \ref{H1.fits}
and \ref{ZEUS.fits} we compare our fits to the H1 and ZEUS DIS data.

One important property of the fits is that the overall normalization
of the quark distribution is quite well determined, as represented by
the coefficient $a_q$.  This is not surprising, since the DIS cross
section is dominated by a quark-induced process.  The systematic shift
to lower values of $a_q$ and $a_g$ as $\alpha_\Pomeron$ increases is entirely
due to the fact that the cross section has a factor
$1/x_\Pomeron^{2\alpha_\Pomeron}$ and that the data are in the region
$x_\Pomeron \leq 10^{-2}$.

\begin{figure}
\centerline{\psfig{file=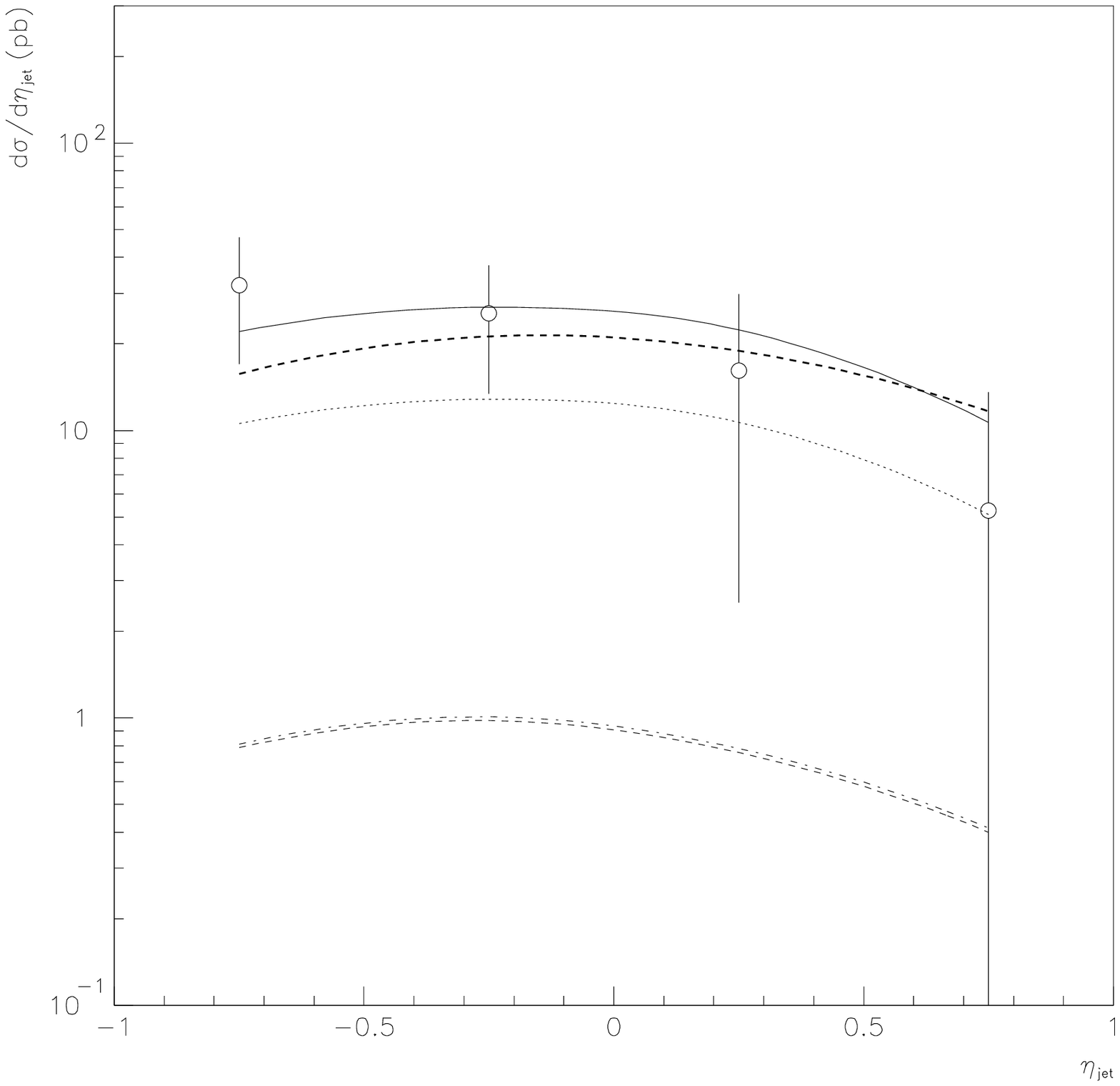,height=4in,clip=}}
\caption{\sf Comparison of the fits for $\alpha_\Pomeron = 1.14$ and the
    ZEUS photoproduction data \protect\cite{ZEUS.photo} used in the fits.
    The code for the lines is the same as in Fig.\
    \protect\ref{H1.fits}.
}
\label{photo.fits}
\end{figure}

The next important property is that a large initial gluon distribution
is strongly preferred.  This can easily be seen from the comparison to
the photoproduction data in Fig.\ \ref{photo.fits}.  With an initial
gluon distribution that is zero,
the cross section (dashed or dot-dashed curve) is over
an order of magnitude below the data.  Even though there are only 4
data points, the improvement when one goes to a parameterization with
a large initial gluon distribution is the dominant effect in
determining the gluon.
The preference is also seen strongly in the $\chi^2$ values for the H1
DIS data --- see Table \ref{Chi2.Table}.  However this preference is also
associated with a negative soft-quark term in the initial parton
densities, which would appear to be unphysical.
We comment on this below.

The relative size of the gluon distributions can be seen
in Table \ref{mom.sums}, which gives the momentum sums for the initial
parton distributions for each of the fits (in the case that
$\alpha_\Pomeron = 1.14$).  
Note that the total momentum sum, as opposed to its quark and gluon
components, is invariant under DGLAP evolution.

\begin{table}
   \begin{center}
   \begin{tabular}{|l!{\vrule width 2pt}*{3}{D{.}{.}{3}|}}
   \hline
        Fit & \multicolumn{1}{c|}{Quarks}
                          & \multicolumn{1}{c|}{Gluon}
                                      & \multicolumn{1}{c|}{Total } \\
   \hline
        A   &     0.160   &   0       &   0.160 \\
        B   &     0.159   &   0.750   &   0.909\\
        C   &     0.156   &   0       &   0.156 \\
        D   &     0.133   &   1.622   &   1.755\\
        SG  &     0.150   &   0.375   &   0.525\\
   \hline
   \end{tabular}
   \end{center}
\caption{\sf Momentum sums
   $\int _{0}^{1} d\beta  \beta  f_{a/\Pomeron}(\beta )$
   at $Q = Q_0 = 2 \, {\rm GeV}$ for the fits with
   $\alpha_\Pomeron = 1.14$.  The quark column represents a sum over
   the 4 light flavors $u$, $d$, $\bar u$, and $\bar d$.
}
\label{mom.sums}
\end{table}

Now let us examine the fits in turn.

Fit A has no gluons and no soft quark term.  A good fit to
the ZEUS DIS data is obtained: $\chi^2/({\rm data\ point})$ is about $10/22$
for the rapidity gap data and $2/3$ to $5/3$ for the LPS data
(depending on the value of $\alpha_\Pomeron$).  However, only a
moderately good fit is obtained for the H1 data:
$\chi^2/({\rm data\ point}) \simeq 70/48$.
The LPS data show a
mild preference for a small value of $\alpha_\Pomeron$, but this
tendency is overwhelmed in the $\chi^2$ by a strong preference of the
H1 data for a larger value: $\alpha_\Pomeron = 1.14$ gives a much
better $\chi^2$ than $\alpha_\Pomeron=1.08$.  However, the
photoproduction data are not reproduced at all.

\begin{figure}
\centerline{\psfig{file=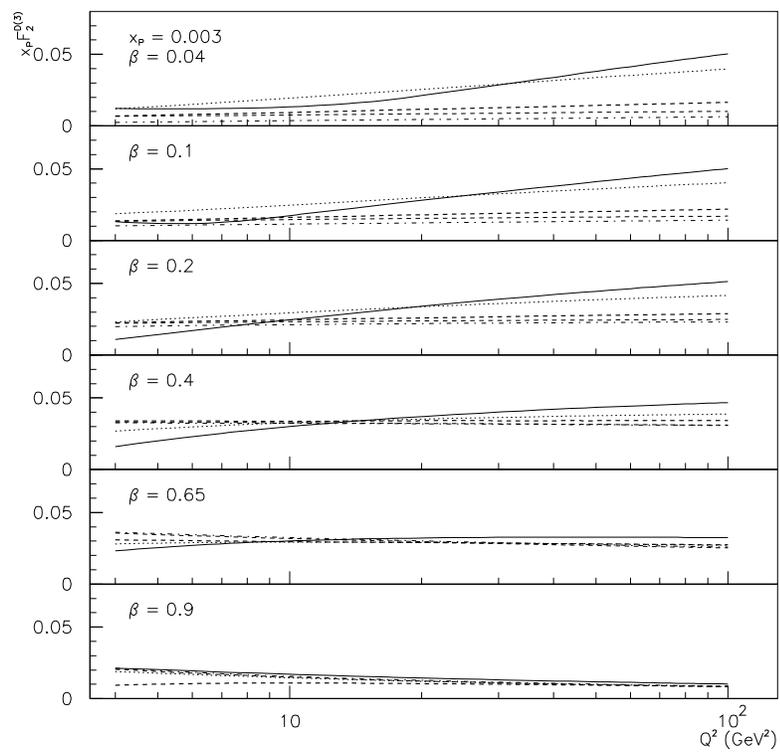,height=4.5in,clip=}}
\bigskip
\caption{\sf $Q^2$ dependence of the fits with $\alpha_\Pomeron = 1.14$.
    The code for the lines is the same as in Fig.\
    \protect\ref{H1.fits}.
}
\label{Q2.dep}
\end{figure}

Fit B differs from fit A by allowing an initial gluon
distribution.
Not surprisingly, this allows us to fit the photoproduction data much better
with the $\chi^2/({\rm data\ point})$ ranging from 2/4 to 4/4.
However, these good $\chi^2$ values are mainly due to to the large errors
in the last two data points.
The $Q^2$ dependence of the diffractive
structure functions, as shown in Fig.\ \ref{Q2.dep}, illustrates the
strong influence of the gluon density on the evolution.
We do not get a particularly good fit to the H1 data.

\begin{figure}
\centerline{\psfig{file=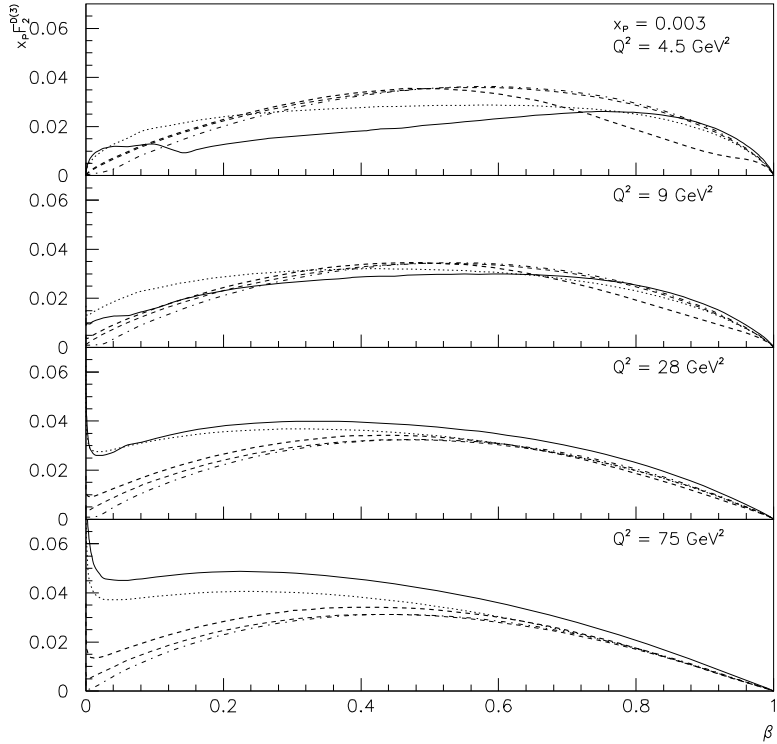,height=4.5in,clip=}}
\bigskip
\caption{\sf $\beta$ dependence of the fits with $\alpha_\Pomeron = 1.14$.
    The code for the lines is the same as in Fig.\
    \protect\ref{H1.fits}.
}
\label{beta.dep}
\end{figure}

We next examine the effect of a soft quark term, in
fit D.  Although, in general, Regge theory suggests that there should
be such a term in the parton densities at some level, what is
surprising is that its coefficient is negative.  The result is in fact
the best of all our fits, including an excellent fit to the
photoproduction data.  The negative soft-quark term cannot represent
the whole story, since it makes the initial quark densities negative
at small $\beta$.  
Notice that the quark distribution only becomes negative when $\beta$
is below the region where we are fitting data, so that we do not have
an unphysical quark density.  
If one wishes to extrapolate our parton densities to low
$\beta$ it would be sensible to replace the initial quark density by
zero whenever the formula gives a negative value.  This is in fact
done automatically by the CTEQ evolution code that we are using, and
one result of this can be seen in Fig.\ \ref{beta.dep}.  In the 
curve for fit D at $Q^2=4.5\,{\rm GeV}^2$, there is a kink a little
below $\beta=0.2$.  Notice that this kink disappears at higher $Q^2$,
when the effects of evolution give a larger positive contribution to
the quark density at small $\beta$.
It is interesting that the restricted set of data to
which we fit provides no significant hint of a soft-quark term if we
restrict to parton densities with no initial glue --- as is seen in
fit C.

\begin{figure}
\centerline{\psfig{file=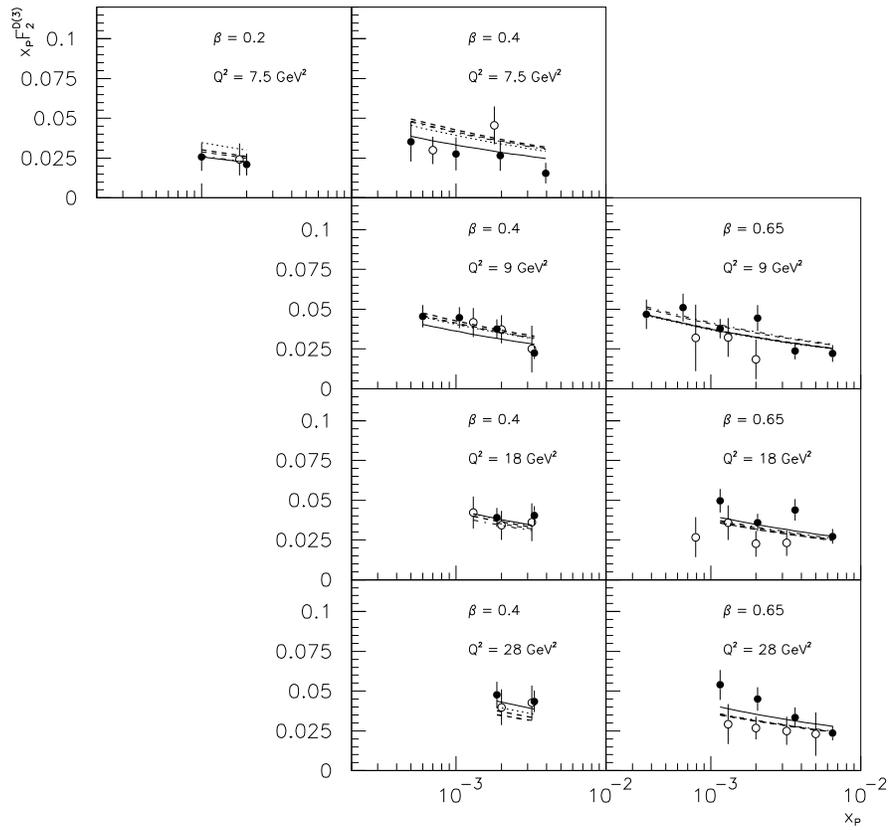,height=5in,clip=}}
\bigskip
\caption{\sf Comparison of the fits with the DIS data from both the H1
    experiment (closed circles) and the ZEUS experiment (open
    circles), in the region where both experiments have data.
    The code for the lines is the same as in Fig.\
    \protect\ref{H1.fits}.
}
\label{combo}
\end{figure}

However, we are not sure to what extent the significance of this
estimate of the soft-quark term is to be taken literally.  If there
were a systematic shift of the data with a series of points moving in a
correlated way by about 1 standard deviation, the soft-quark term
could be much reduced.  Evidence that such a shift is possible is
shown in Fig.\ \ref{combo}.  There we plot DIS data from both
experiments.  Generally the experiments are compatible, but there is a
tendency for the ZEUS data to be about one standard deviation lower
than the H1 data for all the plots at $\beta = 0.65$.  This would have
a significant effect on the $\chi^2$: at the level of ten units, given
the number of points.

Our final fits, SG, have an initial gluon density that is peaked at
large $\beta$, to mimic the one in the fits presented by H1
\cite{H1.DIS}.  Interestingly, we get a good fit to all {\em but} the
H1 data.  It should be remembered, however, that we have found it
appropriate to fit only to a subset of the data, as explained above,
in Sec.\ \ref{sec:selection}.

Finally, we comment on the value of $\alpha_\Pomeron$.  We find that
we prefer the value 1.14 for Fits A, B, C, and SG.  However, Fit D
gives a value of 1.19.  These values are certainly larger than the
value for a soft Pomeron, and the lowest $\chi^2$ is given by fit D
with $\alpha_\Pomeron=1.19$, which is compatible with the value
preferred by H1 \cite{H1.DIS}: $\alpha_\Pomeron = 1.203 \pm 0.020\,
{\rm (stat.)} \pm 0.013 \, {(\rm sys.)} {+0.030 \atop -0.035} \, {\rm
(model)}$.  However, observe that fit D with $\alpha_\Pomeron=1.14$
provides a perfectly adequate fit: $\chi^2=56.3$ for 73 degrees of
freedom, and that the preference for the higher value of
$\alpha_\Pomeron$ is entirely given by the H1 data.

\begin{figure}
\centerline{\psfig{file=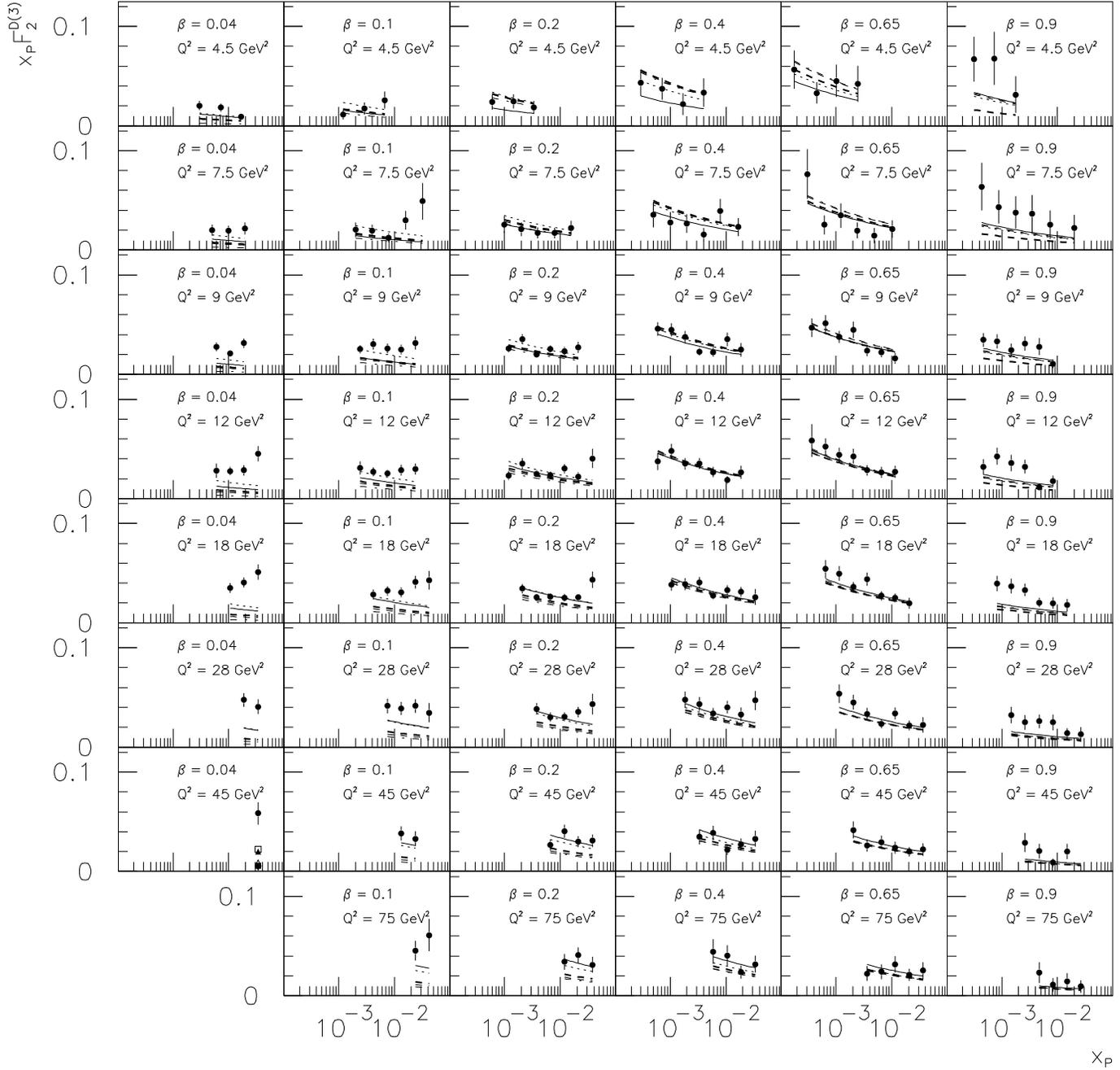,height=8in,clip=}}
\caption{\sf Comparison of the fits for $\alpha_\Pomeron = 1.14$ and all the
    DIS data from H1 \protect\cite{H1.DIS}.
    The code for the lines is the same as in Fig.\
    \protect\ref{H1.fits}.
}
\label{allH1}
\end{figure}

In this context, it is worth examining Fig.\ \ref{allH1}, where we
compare all the H1 data with the predictions of our fits extended
beyond the range where we make the fits.  Some of the data are in a
region of larger $x_\Pomeron$ where we decided that the cross section
is not Pomeron-dominated.  The motivation for excluding certain
regions of
data can be seen particularly clearly in the graphs for $\beta=0.2$.
Furthermore at $\beta=0.9$, the data appear to rise more steeply at
small $x_\Pomeron$ than our fits.  While this is not conclusive, it
suggests that a larger value of the Pomeron intercept,
$\alpha_\Pomeron$, would be needed to fit this subset of the data.  As
we explained in Sec.\ \ref{sec:selection}, these data are in the
resonance region and it is thus not correct to include them in our
fitting or to apply the factorization theorem in this region.

Moreover, it has been established that the Pomeron trajectory is not
universal, since the value of $\alpha_\Pomeron$ in hard scattering is
not the same as in soft scattering.  The proof of factorization
certainly does not require such universality.
Therefore, there is no
reason to assume that the same value of $\alpha_\Pomeron$ applies to
exclusive deep-inelastic processes and to the normal DIS region to
which the factorization formula applies.

Factorization does apply two constraints, however.
First, parton densities are
universal within the class of processes to which the factorization
theorem applies, $\alpha_\Pomeron$ must be the same in these different
processes.  The second constraint arises from DGLAP evolution.
Since evolution relates parton densities at different values of
$Q$ and $\beta$, variations of $\alpha_\Pomeron$ with
$\beta$ and $Q$ cannot be totally arbitrary.
For example, suppose
that at some particular value of $Q$, the value
of $\alpha_\Pomeron$ were larger at large $\beta$ than at small
$\beta$.  Then evolution to larger $Q$ would make the largest value of
$\alpha_\Pomeron$ dominate at all $\beta$.

\subsection{Shape of diffractive parton densities}

Since there are DIS data for a range of values of $\beta$, the data do
provide information on the shape of the diffractive quark
distribution.  For example, we are able to obtain significant
information on the soft quark parameter $\tilde a_q$ in Eq.\
(\ref{paramABCD}).

However, we do not yet have similar information on the shape of the
diffractive gluon distribution.  We have two fits D and SG that
provide good fits to the photoproduction data, but with dramatically
different shapes.  A direct measurement of the shape of the
diffractive gluon distribution can be made in diffractive
photoproduction of dijets
by using
the cross section $d\sigma/d\beta^{\rm OBS}$, where $\beta^{\rm OBS}$
is the longitudinal (light-front) momentum of
the jet pair relative to the Pomeron.  In the leading-order parton
approximation, $\beta^{\rm OBS}$ is exactly the momentum fraction of
the parton in the Pomeron
initiating the hard scattering.

\begin{figure}
\centerline{\psfig{file=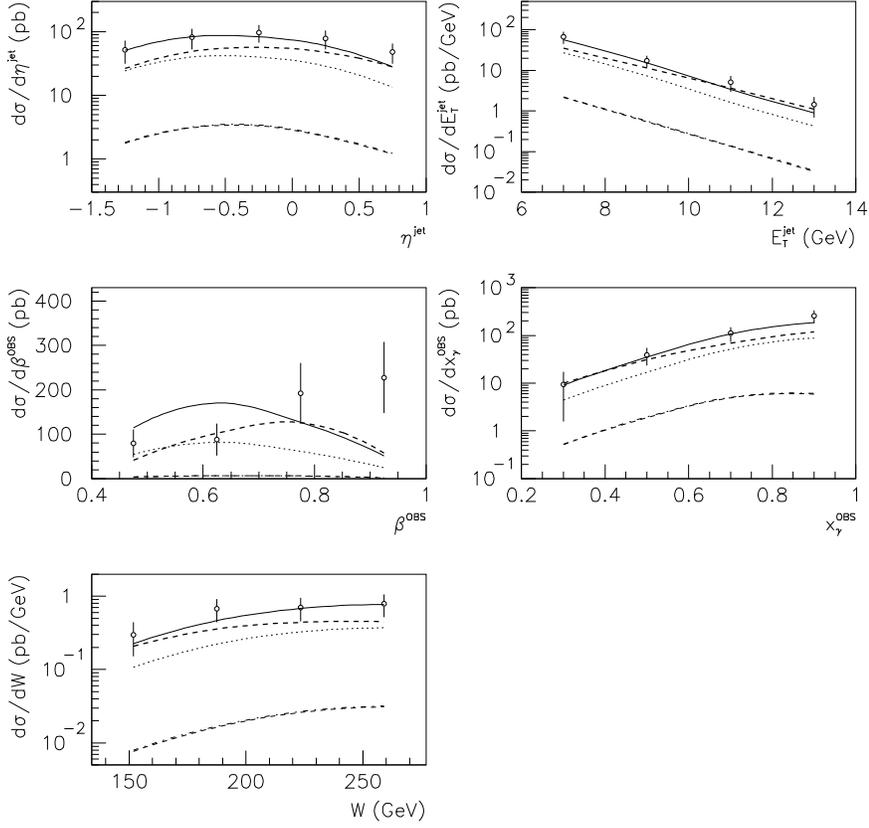,height=5in,clip=}}
\caption{\sf Comparison of the fits for $\alpha_\Pomeron = 1.14$ and the
    preliminary 1994
    ZEUS photoproduction data \protect\cite{ZEUS.Jerusalem}.
    The code for the lines is the same as in Fig.\
    \protect\ref{H1.fits}.
    A double dissociation contribution of $(31\pm13)\%$ has been
    subtracted from the data.  Note that these data were not used in
    the fits.
}
\label{photo.1994}
\end{figure}

We see the implications of these observations in Fig.\
\ref{photo.1994}, where we superimpose our predictions on
preliminary data \cite{ZEUS.Jerusalem} for the diffractive
photoproduction of dijets as a function of each of several kinematic
variables.  The only plot that enables us to distinguish the D and SG
fits is that of the $\beta^{\rm OBS}$ dependence.  The singular gluon
is evidently preferred.  In the other plots, both D and SG reproduce
the normalization and shape of the cross section.

In this same paper \cite{ZEUS.Jerusalem}, fits of diffractive parton densities
were presented.  These were made independently of ours, with the same
kinds of parameterization, but with the inclusion of only the ZEUS
data; the results, particularly as regards the gluon distribution, are
in general agreement with ours.

The preference for a singular gluon distribution is in agreement with
the H1 conclusions \cite{H1.DIS}.  It is interesting that in our fits,
the subset of the H1 data that we use shows the opposite
preference --- see Table \ref{Chi2.Table}.

\section{Kinematics and Cross Sections for Hadron-Induced Processes}
\label{sec:kin}

We now consider the production of $W$ and $Z$ bosons and of jets in
diffractive $p{\bar p}$ collisions.
In addition, we consider $W$ production with explicit
measurement of the distribution of the final-state leptons.
Schematically, these processes are
\begin{eqnarray}
p(p_{1})+{\bar p}(p_{2})&\to & (W\ {\rm or}\ Z) + {\bar p} + X, \nonumber \\
p(p_{1})+{\bar p}(p_{2})&\to & {\rm jet} + {\bar p} + X \nonumber \\
p(p_{1})+{\bar p}(p_{2})&\to & (W\to l + \nu ) + {\bar p} + X.
\label{gnricpro}
\end{eqnarray}
We take the Pomeron to be emitted from the antiproton and the positive
$z$-axis to be along the antiproton's direction.

\subsection{Diffractive jet production}

Consider the diffractive cross section for the production of a
jet with rapidity $y$, in a hadron-hadron collision.
We will assume hard-scattering factorization
\cite{ISorig,CTEQpom}.
At leading order (LO), the
hard-scattering process is $2 \to  2$ at the parton level, and
results in a cross section of the form
\begin{equation}
{d\sigma ^{\rm jet} \over dy} =
\sum _{a,b}\int dE_{T}~2E_{T}
\int dy' \int dx_{\Pomeron}
\, f_{{\Pomeron}/{\bar p}}(x_{\Pomeron},\mu )
\,f_{a/p}(x_{a},\mu )
\,f_{b/{\Pomeron}}(x_{b},\mu )x_{a}x_{b}
\frac {d \hat{\sigma }^{\rm jet}_{ab}}{d \hat t} ,
\label{diffcs1}
\end{equation}
where the sum is over all the active parton (quark, antiquark and gluon)
flavors.  The integration variables are $E_{T}$, the transverse
energy of the jet, $y'$, the rapidity of the other jet,
and $x_{\Pomeron}$, the momentum fraction of the Pomeron.  The
momentum fractions of the partons, relative to their parent
proton and Pomeron are
\begin{eqnarray}
x_{a} = {E_{T} \over \sqrt {s}}(e^{-y}+e^{-y'}) ~~\mbox{and}~~
x_{b} = {E_{T} \over \sqrt {s}x_{\Pomeron}}(e^{y}+e^{y'}).
\label{xpartons}
\end{eqnarray}
The functions $f_{a/p}(x_{a})$ and $f_{b/{\Pomeron}}(x_{b})$
are the number densities
of partons in the proton  and Pomeron,
respectively,
while $f_{{\Pomeron}/{\bar p}}$ is the same Pomeron flux factor that
we used in Sec.\ \ref{sec:DIS} [see Eq.~(\ref{DLflux})].
$d\hat{\sigma }^{\rm jet}_{ab}/d\hat{t}$ is the
partonic hard-scattering coefficient, and $\mu $ is the factorization
scale, which we set equal to $E_{T}$.
The limits on the integrations are determined by the experimental
conditions.

The diffractive cross section given by Eq.~(\ref{diffcs1}) has
the same structure as the factorized form of the corresponding
inclusive cross section (i.e., without the diffractive requirement),
except for the Pomeron flux factor
and the parton densities in the Pomeron.
The same hard-scattering coefficient and
nucleon parton distribution functions appear in both cross sections.

The cross section given by Eq.~(\ref{diffcs1}) has contributions
from a range of subprocesses.  The indices $a,b$ labeling the
incoming partons range over the gluon and all the flavors of
quarks and antiquarks. The
leading order (LO) form of the partonic cross
section $d\hat{\sigma }^{\rm jet}_{ab}/d\hat{t}$ may be found in
\cite{eichetal}.

\subsection{Diffractive $W$ and $Z$ production}

The cross section for the diffractive production of weak vector bosons
is given by
\begin{equation}
  \sigma ^{VB} = \sigma_0^{VB}
    \sum _{a,b} \int {dx_{\Pomeron}\over x_{\Pomeron}}
    \int {dx_b\over x_b} \int {dx_a\over x_a}
    \, f_{{\Pomeron}/{\bar p}}(x_{\Pomeron})
    \, f_{b/{\Pomeron}}(x_{b})
    \, f_{a/p}(x_{a})
    \, {\tilde C_{ab}^{VB}}
    ~ \omega_{ab}\biggl({\tau\over x_ax_bx_{\Pomeron}},\alpha_s\biggr),
\label{vbincldfcs}
\end{equation}
where
$\sigma_0^{VB}={\sqrt {2}\pi G_{F}M_{VB}^{2} / 3s}$,
$M_{VB}=M_W \ {\rm or} \ M_Z$ is the vector boson mass,
$G_{F}$ is the Fermi constant,
$x_b,x_a$ are momentum fractions of partons from the Pomeron and proton,
respectively, and $\tau=M_{VB}^2/s$.
For $W$ bosons, ${\tilde C^W_{qq'}}=|V_{qq'}|^2$ if
$e_q+e_{q'}=\pm 1$ and zero otherwise, where $q$ denotes a quark flavor,
$e_q$ the fractional charge of quark $q$ and
$V_{qq'}$ is the Cabibbo-Kobayashi-Maskawa matrix element.
For $Z$ bosons,
${\tilde C}^Z_{q{\bar q}}=1/2-2|e_q|\sin^{2}\theta _{W}
    +4|e_q|^{2}\sin^{4}\theta _{W}$,
where $\theta _{W}$ is the Weinberg or weak-mixing angle.
Similar expressions apply for ${\tilde C^W_{qg}}$ and ${\tilde C}^Z_{qg}$
which are relevant for gluon-induced scattering.
The hard-scattering function $\omega_{ab}$ in the $\overline{{\rm MS}}$ scheme
and to next-to-leading order (NLO)
in the QCD strong coupling $\alpha_s$
can be found in \cite{dyhardpart}.

\subsection{Diffractive production of leptons from the $W$}

Since leptonic decays of $W$ bosons include an unobserved
neutrino, it is useful to compute the distribution of the
observed charged lepton.
The general formula for the distribution of leptons from $W$
production has the
same form as that for jet production, Eq.~(\ref{diffcs1}).
In this case, we are only going to compute cross sections at leading order.
Data have not been published for this particular cross section but
since it is directly measurable, we think it is a useful quantity to work with.

For the specific process
$p+{\bar p}\to (W^{-}\to e + {\bar \nu }_{e})+{\bar p}+X$, we have
the leading-order cross section for quark-antiquark annihilation to a
lepton pair:
\begin{equation}
{d\hat{\sigma }^{\rm lep}_{ab}\over d\hat{t}}\simeq
{G_{F}^{2}\over 6M_{W}\Gamma _{W}} \, \tilde {C}^W_{ab}
\, \delta (x_{a}x_{b}s-M_{W}^{2})
\, \hat{u}^{2},
\label{wmpartcs}
\end{equation}
where $\Gamma _{W}$ is the width of the $W$ boson and
$\hat{u}=-x_{b}x_{\Pomeron}\sqrt {s}E_{T}e^{-y}.$
Using Eq.~(\ref{wmpartcs}) in Eq.~(\ref{diffcs1}), one obtains
the following cross section at the hadronic level:
\begin{equation}
    {d\sigma ^{\rm lep} \over dy}=
    \sum _{a,b}\int {dx_{\Pomeron}\over x_{\Pomeron}}
    \int dE_{T}
    \, f_{{\Pomeron}/{\bar p}}(x_{\Pomeron})
    \, f_{b/{\Pomeron}}(x_{b})
    \, f_{a/p}(x_{a})
    \, \tilde {C}_{ab}^W
    \, \left[{\hat{u}^{2}G_{F}^{2}\over 6s\Gamma _{W}
          [(M_{W}/2E_{T})^{2}-1]^{1/2}}
    \right],
\label{wmdiffcs}
\end{equation}
where
$x_{a}$ and $x_{b}$ are now given by
\begin{eqnarray}
x_{a}&=& \frac {M_{W} e^{-y}}{\sqrt s}
     \left[\frac {M_{W}}{2E_{T}}
        + \sqrt { \left(\frac {M_{W}}{2E_{T}}\right)^{2}-1 }
     \right],
\nonumber \\
x_{b}&=&{M_{W}^{2}\over s}{1\over x_{a}x_{\Pomeron}}.
\label{wmxpart}
\end{eqnarray}
We have suppressed the scale dependence of
the functions $f_{i/j}$ in Eqs.~(\ref{vbincldfcs})
and (\ref{wmdiffcs}); in actual computations, we set the scale
equal to the vector boson mass.
A similar equation may be
obtained for the $W^{+}$ cross section.

\subsection{Inclusive cross sections}

Since we are particularly interested in the percentage of events
that are diffractive, we also need to calculate the inclusive
cross sections, that is, the ones without the diffractive
requirement on the final state.
The analog to Eq.~(\ref{diffcs1}) for the
inclusive cross section for jet production is the
standard formula
\begin{equation}
    {d\sigma ^{\rm jet,\ incl} \over dy}=
    \sum _{a,b}\int dE_{T}2E_{T}
    \int dy'
    \, f_{a/p}(x_{a},\mu ) \, f_{b/{\bar p}}(x_{b},\mu )x_{a}x_{b}
    \frac {d\hat{\sigma }^{\rm jet}_{ab}}{d \hat t},
\label{nondiffcs1}
\end{equation}
where $x_{a}$ is given in Eq.~(\ref{xpartons}), while $x_{b}$ is now
$x_{b}=(e^{y}+e^{y'}) E_{T}/ \sqrt {s}$.

For the leptons from
$W^{-}$ production, the inclusive version of Eq.~(\ref{wmdiffcs}) is
\begin{equation}
{d\sigma ^{\rm lep,\ incl} \over dy} =\sum _{a,b}
\int dE_{T}
\, f_{b/{\bar p}}(x_{b},\mu )f_{a/p}(x_{a},\mu )
\, \tilde {C}_{ab}^W
\,
\left[{\hat{u}^{2}G_{F}^{2}\over 6s\Gamma _{W}[(M_{W}/2E_{T})^{2}-1]^{1/2}}
\right],
\label{wmndiffcs}
\end{equation}
with a similar equation for $W^{+}$ production.  In Eq.~(\ref{wmndiffcs}),
$\hat{u}=-x_{b}\sqrt {s}E_{T}e^{-y}$,
$x_{a}$ is as defined in Eq.~(\ref{wmxpart}) while $x_{b}$ is now given by
$x_{b}=M_{W}^{2}/x_{a}s$.

The analog to Eq.~(\ref{vbincldfcs}) for the inclusive total cross section
for vector boson production is
\begin{equation}
  \sigma ^{\rm VB}=\sigma_0^{\rm VB} \sum _{a,b}
    \int {dx_a\over x_a} \int {dx_b\over x_b}
    \, f_{a/p}(x_{a}) \, f_{b/{\bar p}}(x_{b})
    \, \tilde{C}^{\rm VB}_{ab}
    ~ \omega_{ab}\left({\tau\over x_ax_b},\alpha_s\right).
\label{vbinclndfcs}
\end{equation}
%

\section{Numerical Calculations of $W$ and $Z$ Production}
\label{sec:VB.calcs}

For the calculations in this section, the factorization scale in the parton
distributions was set to $M_{VB}$.  The values of the electroweak parameters
that appear in the various formulae were taken from the particle data
handbook \cite{PDG},
and we use only four flavors ($u$, $d$, $s$, $c$)
in the weak mixing matrix,
with the Cabibbo angle $\theta _{C} = 0.2269$.

\subsection{Comparison with previous calculations}

Bruni and Ingelman \cite{BI} computed diffractive $W/Z$ cross
sections
neglecting any $Q^{2}$ evolution of the parton distributions in
the Pomeron.  At $\sqrt {s}=1800 \ {\rm GeV}$, they obtained the
following
diffractive fractions ($R=\sigma ^{\rm diff} /\sigma ^{\rm incl}$):
$R_{W^{+}+W^{-}}\simeq 20\%$ and
$R_{Z}\simeq 17\%$ for total $W$ and $Z$ production, respectively.
These rates are substantially larger than the few percent measured by
CDF in \cite{CDF.wgap}.

As we will now explain, when one uses evolved Pomeron parton densities
from our fits to data from HERA,
one obtains substantially smaller rates than the Bruni-Ingelman ones.
To understand these small rates, we first
verify that we can reproduce the Bruni-Ingelman results.
For these, we used
their unevolved hard quark distribution in
the Pomeron (given by their Eq.~(4)),
the same cut on $x_{{\Pomeron}}$: $x_{{\Pomeron}}^{\rm max}=0.1$, the
EHLQ1 parton distributions in the proton
and the Ingelman-Schlein (IS) flux factor:\footnote{
    Note that since our purpose in using the IS flux is to
    compare our results with the Bruni-Ingelman calculations, we
    have used a Pomeron intercept of unity instead of the more
    accurate value used in our fits.
}
\begin{equation}
f_{{\Pomeron}/p}^{\rm IS}(x_{{\Pomeron}})=\int dt {1\over 2.3x_{{\Pomeron}}}
\biggl(3.19e^{8t}+0.212e^{3t}\biggr).
\label{ISflux}
\end{equation}
The flux in Eq.~(\ref{ISflux}) differs by a factor of 1/2 from that in
\cite{BI} because here we consider the case when only $\bar p$ diffracts
while \cite{BI} considers the case when either $p$ or $\bar p$ diffracts.

Next, we evolved their Pomeron parton distributions and
recalculated the
cross sections.
Finally, to provide our best estimates of the rates,
we repeated the calculations using CTEQ4M for
the parton densities in
the proton/antiproton and using our fits
for the parton densities in the Pomeron, all with proper
evolution.
The cross sections were calculated using
Eqs.~(\ref{vbincldfcs}) and (\ref{vbinclndfcs})
and the results obtained are summarized in Tables
\ref{table:inclWZ} to \ref{table:diffWZ0.01}.

First, in Table \ref{table:inclWZ}, we show
the {\em inclusive}
cross section, $\sigma ^{\rm incl}$,
which will give the
denominator for the fraction of the cross section that is
diffractive.
We present the leading order (LO) result from \cite{BI} as well as our
leading and next-to-leading order (NLO) results.
At leading order, one observes that the
use of the more up-to-date CTEQ4M densities in the proton instead
of the EHLQ1 densities used by Bruni and Ingelman leads to cross sections that
are 20\% to 30\% higher.  Including the next-to-leading order contributions
leads to another similar increase in the cross sections.

\begin{table}
\begin{center}
   \begin{tabular}{|c!{\vrule width 2pt}c!{\vrule width
                   2pt}r|r!{\vrule width 2pt}r|r|} \hline
           & \multicolumn{1}{c!{\vrule width 2pt}} {EHLQ1} &
\multicolumn{2}{c!{\vrule width 2pt}} {EHLQ1} &
\multicolumn{2}{c|} {CTEQ4M} \\ \cline{2-6}
           & Ref.\cite{BI} LO & \multicolumn{1}{c|} {LO} &
\multicolumn{1}{c!{\vrule width 2pt}} {NLO} & \multicolumn{1}{c|} {LO} &
\multicolumn{1}{c|} {NLO} \\ \hline
   $W^{+}+W^{-}$ & 14000 & 14300 & 18100 & 18700 & 23500   \\ \hline
   $Z$     &  4400 &  4400 & 5600 &  5500 & 6900 \\ \hline
   \end{tabular}
\end{center}
\caption{\sf Inclusive cross sections $\sigma ^{W,Z\,\rm incl}$ (pb) for
         weak vector boson production.
}
\label{table:inclWZ}
\end{table}

The diffractive cross sections $\sigma ^{W,Z\, \rm diff}$ are
shown in Tables
\ref{table:diffWZ0.1} and \ref{table:diffWZ0.01}.
In the columns labeled ``BI'', we used the Bruni-Ingelman parton
density in the Pomeron and the EHLQ1 parton densities in the
proton,
together with the Ingelman-Schlein flux factor (\ref{ISflux}).
In the other columns we used our fits for the parton densities in
the Pomeron together with the CTEQ4M parton distributions in the
proton;
we use the Donnachie-Landshoff form for the flux factor,
(\ref{DLflux}), and $\alpha_\Pomeron=1.14$.
First,
we use the same cut $x_{{\Pomeron}}^{\rm max} = 0.1$ as was used by
Bruni and Ingelman, to produce Table \ref{table:diffWZ0.1}.
However, this allows $x_{\Pomeron}$ to be rather larger
than where Pomeron exchange is expected to dominate.  So we also
made calculations with $x_{{\Pomeron}}^{\rm max} = 0.01$,
for which the results
are shown in Table \ref{table:diffWZ0.01}.

In column 3 of Table \ref{table:diffWZ0.1} we show our leading order results
when we use the same unevolved parton densities as Bruni and
Ingelman; we agree with their cross sections (column 2).
Then we repeat the calculations but with
correctly evolved parton densities in the Pomeron, with the
Bruni-Ingelman formula being used as the initial data for the
evolution at $Q_{0}^{2}=4 \, {\rm GeV}^{2}$ (column 5).  The corresponding
next-to-leading order cross sections are shown in columns 4 and 6.
We see that at either LO or NLO, evolution of the Pomeron parton densities
leads to about a 30\% reduction in the cross section.

\begin{table}
\begin{center}
   \begin{tabular}{|c!{\vrule width 2pt}c!{\vrule width
                   2pt}r|r!{\vrule width 2pt}r|r!{\vrule width
                   2pt}r|r!{\vrule width 2pt}r|r|} \hline
   Pomeron: &  BI\cite{BI} & \multicolumn{2}{c!{\vrule width 2pt}} {BI}
   & \multicolumn{2}{c!{\vrule width 2pt}} {BI}
   & \multicolumn{2}{c!{\vrule width 2pt}} {Fit A}
   & \multicolumn{2}{c|} {Fit D} \\
   &  unevolved & \multicolumn{2}{c!{\vrule width 2pt}} {unevolved}
   & \multicolumn{2}{c!{\vrule width 2pt}} {evolved}
   & \multicolumn{2}{c!{\vrule width 2pt}} {evolved}
   & \multicolumn{2}{c|} {evolved} \\ \cline {2-10}
   Proton: & EHLQ1 & \multicolumn{2}{c!{\vrule width 2pt}} {EHLQ1}
   & \multicolumn{2}{c!{\vrule width 2pt}} {EHLQ1}
   & \multicolumn{2}{c!{\vrule width 2pt}} {CTEQ4M}
   & \multicolumn{2}{c|} {CTEQ4M}  \\ \cline {2-10}
  & LO & \multicolumn{1}{c|} {LO} & \multicolumn{1}{c!{\vrule width 2pt}} {NLO}
  & \multicolumn{1}{c|} {LO} & \multicolumn{1}{c!{\vrule width 2pt}} {NLO}
  & \multicolumn{1}{c|} {LO} & \multicolumn{1}{c!{\vrule width 2pt}} {NLO}
  & \multicolumn{1}{c|} {LO} & \multicolumn{1}{c|}{NLO}
   \\ \hline
$W^{+}+W^{-}$ & 1400 & 1400 & 1800 & 1000 & 1300 & 300 & 390 & 650  & 810
\\ \hline
$Z$ & 380 & 380 & 480 & 260 & 330 &  77 & 100 & 170 & 210 \\ \hline
\end{tabular}
\end{center}
\caption{\sf Diffractive cross section $\sigma ^{W,Z\, \rm diff} $ (pb)
    for weak vector boson production when only $\bar p$ diffracts and
    with $x_{\Pomeron}^{\rm max} = 0.1$.
    The cross sections using the BI distributions were computed
    with $\alpha_\Pomeron=1$, as in Ref.\ \cite{BI}, but the cross
    sections using fits A and D were computed with
    $\alpha_\Pomeron=1.14$.
}
\label{table:diffWZ0.1}
\end{table}

\begin{table}
\begin{center}
   \begin{tabular}{|c!{\vrule width 2pt}r|r!{\vrule width
                   2pt}r|r!{\vrule width 2pt}r|r|} \hline
Pomeron: & \multicolumn{2}{c!{\vrule width 2pt}} {BI}
   & \multicolumn{2}{c!{\vrule width 2pt}} {Fit A}
   & \multicolumn{2}{c|} {Fit D} \\
   & \multicolumn{2}{c!{\vrule width 2pt}} {evolved}
   & \multicolumn{2}{c!{\vrule width 2pt}} {evolved}
   & \multicolumn{2}{c|} {evolved} \\ \cline {2-7}
   Proton: & \multicolumn{2}{c!{\vrule width 2pt}} {EHLQ1}
   & \multicolumn{2}{c!{\vrule width 2pt}} {CTEQ4M}
   & \multicolumn{2}{c|} {CTEQ4M}  \\ \cline {2-7}
  & \multicolumn{1}{c|} {LO} & \multicolumn{1}{c!{\vrule width 2pt}} {NLO}
  & \multicolumn{1}{c|} {LO} & \multicolumn{1}{c!{\vrule width 2pt}} {NLO}
  & \multicolumn{1}{c|} {LO} & \multicolumn{1}{c|}{NLO}
   \\ \hline
$W^{+}+W^{-}$ & 25 & 38 & 9 & 14 & 14 & 21 \\ \hline
$Z$ & 3.2 & 5.0& 1.1 & 1.8 & 1.6 & 2.5 \\ \hline
\end{tabular}
\end{center}
\caption{\sf Diffractive cross section $\sigma ^{W,Z\, \rm diff} $ (pb)
    for weak vector boson production when only $\bar p$ diffracts,
    but now with $x_{\Pomeron}^{\rm max} = 0.01$.
    The cross sections using the BI distributions were computed
    with $\alpha_\Pomeron=1$, as in Ref.\ \cite{BI}, but the cross
    sections using fits A and D were computed with
    $\alpha_\Pomeron=1.14$.
}
\label{table:diffWZ0.01}
\end{table}

We also present in Table~\ref{table:fractions} the diffractive fractions for
total $W$ production when either $p$ or $\bar p$ diffracts.  The fractions
are obtained by dividing twice the diffractive cross sections in
Tables~\ref{table:diffWZ0.1} and \ref{table:diffWZ0.01} (which are for
{\it single-sided} diffraction) by the appropriate inclusive cross section
in Table~\ref{table:inclWZ}.

The diffractive fraction obtained from the evolved BI Pomeron
parton distribution, using columns 3 and 4 of Table \ref{table:inclWZ}
for $\sigma ^{W,Z\, \rm incl}$,
is about 14\% for $W$ production, compared with the
20\% that is obtained using the unevolved BI Pomeron
distributions.
The corresponding percentages for $Z$ production are a little
smaller: 12\% (evolved) and 17\% (unevolved).

In the last four columns of Tables \ref{table:diffWZ0.1} and
\ref{table:diffWZ0.01} we present the results when two of our
fits (with $\alpha_{\Pomeron}=1.14$) shown in
section \ref{sec:fits} are used.
Fit A is the one with a simple hard
quark distribution and no glue as the initial values, while fit D,
which has both quarks and gluons initially,
is our best fit overall.
Now fit A does not have the large gluon content that is necessary to
fit the photoproduction data, so cross sections computed using fit A
cannot be said to represent good predictions.  However it is adjusted
to fit DIS data, so that a comparison of predictions using fits A and
D pinpoints situations where the large gluon distribution has a large
effect.

We have also calculated the cross sections resulting when we use
the versions of the diffractive parton densities with a higher
value of the Pomeron intercept, $\alpha_\Pomeron = 1.19$.  We
find that the cross sections are reduced by 10\% to 20\%,
depending on the value of $x_\Pomeron^{\rm max}$.  The reduced
cross section arises because the diffractive parton densities are
constrained to fit ZEUS and H1 data at fairly small values of
$x_\Pomeron$ and we are now calculating cross sections at higher
values of $x_\Pomeron$.  So an increase in $\alpha_\Pomeron$
results in a decrease in our calculated cross section for the
hadron-induced processes.

The LO and NLO cross sections resulting from fit A (columns 7 and 8 of Table
\ref{table:diffWZ0.1}) are only about
30\% of the evolved BI cross sections.
(We will indicate the sources of this difference below, in Sec.\
\ref{Compare.BI}.)
The diffractive fractions obtained from this fit, using
the CTEQ4M entries in Table \ref{table:inclWZ},
are 3.3\% (2.9\%) for $W\ (Z)$ production,
as shown in Table \ref{table:fractions}.

The quark distributions in fit D are about 20\% higher than in fit
A---see the values of $a_q$ in Table \ref{Params}.  However the cross
sections for $W$ production with fit D exceed those with fit A by a
substantially larger factor, particularly at $x_\Pomeron^{\rm max} =
0.1$, where the cross section is more than a factor 2 higher.  This
arises because of evolution: the large gluon distribution in fit D
increases the quark distribution at $\mu=M_W$ compared with the case
without the large gluon distribution.  The increased quark density is
most pronounced at small fractional momentum.  Thus the effect is
larger at $x_\Pomeron^{\rm max} = 0.1$ than at $x_\Pomeron^{\rm max} =
0.01$, since in the first case, the quark from the Pomeron that makes
the $W$ has a smaller fractional momentum relative to the Pomeron.
The NLO contributions further increase the fit D cross sections by
24\%.
Even so, the cross sections are still smaller, by a factor of
1.6, than
the ones from evolved BI Pomeron parton distributions.  The rates
from fit D (using NLO values) are
6.9\% (6.1\%) for $W\ (Z)$ production.
These rates agree with those obtained by Kunszt and
Stirling \cite{KS} with their Model 2 for diffractive
parton distributions.

The data from which our fits were extracted
used a conservative cut
on the Pomeron momentum, $x_{{\Pomeron}}^{\rm max}=0.01$.
The Pomeron flux factor allows for the $x_{\Pomeron}$ dependence, but
to ensure maximum compatibility with the HERA data without the
assumption of standard Regge behavior, the same cut should be
applied to the cross sections in hadron-hadron collisions.  This
results in the cross sections in Table \ref{table:diffWZ0.01},
which therefore represent our most accurate prediction of
diffractive $W$ and $Z$ production, given only the assumption of
hard-scattering factorization,
{\em which of course we wish to test}.
Notice that with this cut the diffractive
cross sections are over an order of magnitude smaller
than with
$x_{{\Pomeron}}^{\rm max}=0.1$.
The percentages obtained with this cut on $x_{{\Pomeron}}$
for $W\ (Z)$ production are 0.12\% (0.05\%) and 0.18\% (0.07\%)
using fits A and D, respectively,
as shown in Table \ref{table:fractions}.
The large reduction is due to
the fact that we are not far from an effective kinematic limit:
the cut on $x_{\Pomeron}$ gives a maximum proton-Pomeron energy of
180 GeV, and partons typically do not carry the whole of the
energy of their parent hadrons.

\begin{table}
\begin{center}
   \begin{tabular}{|c!{\vrule width 2pt}c!{\vrule width
                   2pt}c|c!{\vrule width 2pt}c|c|}\hline
                & BI (unevolved) & Fit A & Fit D & Fit A & Fit D \\
                & $x_{\Pomeron}^{\rm max} = 0.1$
                    & $x_{\Pomeron}^{\rm max} = 0.1$
                        & $x_{\Pomeron}^{\rm max} = 0.1$
                            & $x_{\Pomeron}^{\rm max} = 0.01$
                               & $x_{\Pomeron}^{\rm max} = 0.01$
        \\
        \hline
        $W^{+}+W^{-}$ & 20\% & 3.3\% & 6.9\% & 0.12\% & 0.18\% \\
        \hline
        $Z$     & 17\% & 2.9\% & 6.1\% & 0.05\% & 0.07\% \\ \hline
    \end{tabular}
\end{center}
\caption{\sf Diffractive fractions using NLO cross sections
for $W^++W^-$ and $Z$ production when either $p$ or $\bar p$ diffracts.}
\label{table:fractions}
\end{table}

\subsection{Why are the fractions smaller than from BI?}
\label{Compare.BI}

Although the data used in our fits support a ``hard'' quark
distribution
in the Pomeron, we predict that the diffractive
$W$ and $Z$ cross sections are
much smaller than those predicted by
Bruni and Ingelman, who also used hard quark distributions.
For example, the diffractive fraction for $W$ production computed using fit A,
is six times smaller than Bruni and Ingelman's fraction (see
Table~\ref{table:fractions}).

Since Bruni and Ingelman's work served as an initial benchmark for
subsequent work, it is interesting to understand the sources of this
factor.  We first address fit A, since that is our parameterization
that is closest to Bruni and Ingelman's.  The factor between the
diffractive rates arises as an accumulation of several modest factors:
\begin{itemize}

\item
    A factor 0.9 because of the use of the
    CTEQ4M instead of the obsolete EHLQ1
    distributions in the proton.  (The denominator in the ratio of
    diffractive to inclusive cross sections increases by more than the
    numerator.)

\item
    A factor 0.7 for the effect of the evolution of the parton
    densities in the Pomeron.

\item
    A factor 1.7 for the use of the Donnachie-Landshoff flux
    factor instead of the Ingelman-Schlein flux factor, when the
    momentum sum is kept fixed.  We have found that this factor
    arises from the following:
    \begin{itemize}
        \item
    A factor of 2.5 to allow for our larger value of $\alpha_\Pomeron$.

        \item
    A factor of 0.7 to allow for the effects of the Pomeron slope $\alpha'$.
    \end{itemize}
\item
    A factor 0.16 because the DIS data indicate that the quarks
    have a momentum sum substantially less than the value of unity
    that was assumed by Bruni and Ingelman.  \footnote{
        Note that in the case of the diffractive DIS cross section,
        this small momentum sum is mostly compensated by the effects
        of our larger value of $\alpha_\Pomeron$, which increases the
        cross sections at small $x_\Pomeron$.
    }

\end{itemize}

The first three factors in fact cancel.  So one possible view is that
the smallness of our results compared with those of Bruni and Ingelman
arises essentially because of the change in the momentum sum of the
quarks required by a fit to the data.  An alternative view arises when
one observes that the Pomeron-proton coupling is obtained by fitting
data on high energy scattering, and that if one increases
$\alpha_\Pomeron$, then the value of the Pomeron-proton coupling has
to be decreased to keep the cross section at some particular energy
fixed.  (Of course, the energy dependence of the cross section would
not be fitted so well.)  The variable in high-energy scattering that
corresponds to $1/x_\Pomeron$ for hard diffraction is $s/M^2$.  Now,
the typical value of $x_\Pomeron$ in the data that we fit is about
$10^{-3}$, which corresponds to $s \sim 1000 \,{\rm GeV}^2$, i.e., a
fixed target energy of around $500 \, {\rm GeV}$.  Therefore it is
possible to argue that the factor of 2.5 for the larger value of
$\alpha_\Pomeron$ should be combined with the factor of 0.16 for the
momentum sum, to produce a factor of 0.4 for the momentum sum with the
Pomeron-proton coupling fixed at a value appropriate for fixed-target
CERN and Fermilab experiments.  The overall reduction in our rates
compared with those of Bruni and Ingelman then arises as a product of
several factors, all less than unity.

The effects of this decrease in cross section in going from the
Bruni-Ingelman ansatz to our fitted distribution A are then somewhat
compensated by the effects of the large gluon distribution we find in
fit D.

\subsection{Lepton distributions for $W$ production at the Tevatron}

In this section, we present our results for $W$ production, but now
with cuts on the emitted lepton $l$.  Specifically, we calculate the
electron's (or positron's) rapidity $(y)$ distribution from
Eq.~(\ref{wmdiffcs}) for the diffractive process, and
Eq.~(\ref{wmndiffcs}) for the inclusive one.  For the parton
distributions in the Pomeron, we use our five fits with
$\alpha_{\Pomeron}=1.14$---see
Eqs.~(\ref{paramABCD})--(\ref{paramSG}) and Table
\ref{Params}---evolved up to the $W$ mass.
We imposed
a cut of 20 GeV on the $E_{T}$ of the emitted lepton, and
we integrated $x_{{\Pomeron}}$
up to $x_{{\Pomeron}}^{\rm max}=0.01$.

\begin{figure}
\centerline{\psfig{file=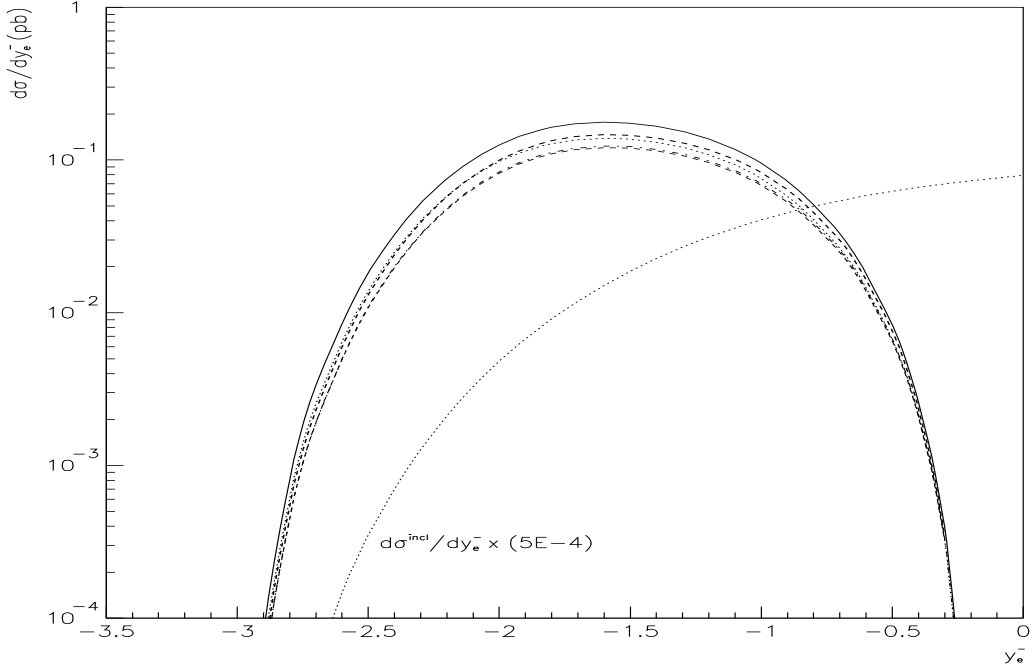,height=4in,width=6in,clip=}}
\caption{\sf Rapidity distribution of $e^{-}$ in $W^{-}$ production with the
        cut $x_{\Pomeron} < 0.01$.  Solid curve results from using fit D,
    upper dashed with fit SG, dotted with fit B, lower dashed with
    fit A and dot-dashed with fit C.  The lower dotted curve is the
    inclusive cross section scaled down by a factor $5\times 10^{-4}$. }
\label{fig1Wm}
\end{figure}

Figure \ref{fig1Wm} shows our results for $W^{-}$ production.
For comparison, we also show the inclusive cross section rescaled by
$5\times 10^{-4}$ as represented by the lower dotted curve.
The diffractive cross sections exhibit a strong fall-off
in the region $y_{e}>-0.2$ that is a consequence of the requirement
of a rapidity gap.  This fall-off is of course not present in the
inclusive cross section.

The diffractive cross sections are about
2\% to 4\% of the inclusive
one at the
left edge
of the plots (at $y=-3$) depending on the fit used.
At about $y=-1.6$ where the diffractive cross sections peak, this fraction
drops to about
0.4\% to 0.6\% of the inclusive cross section.
The cross sections using high glue fits D, B and SG, denoted by solid,
dotted and upper dashed curves, respectively, are
larger (D gives largest cross section)
than those using the low glue fits A (dashed) and C (dot-dashed)
which overlap in the figure.
The differences between the cross sections reflect first the size of
the quark densities and then, in fits B, D and SG, the effects of a
large gluon distribution on the evolution of the quark distribution.
For example, fit SG has a smaller quark distribution than fit A, but
its large gluon distribution pulls the cross section above that given
by fit A.  However the differences are moderate, at most a factor of
1.5.

\begin{figure}
\centerline{\psfig{file=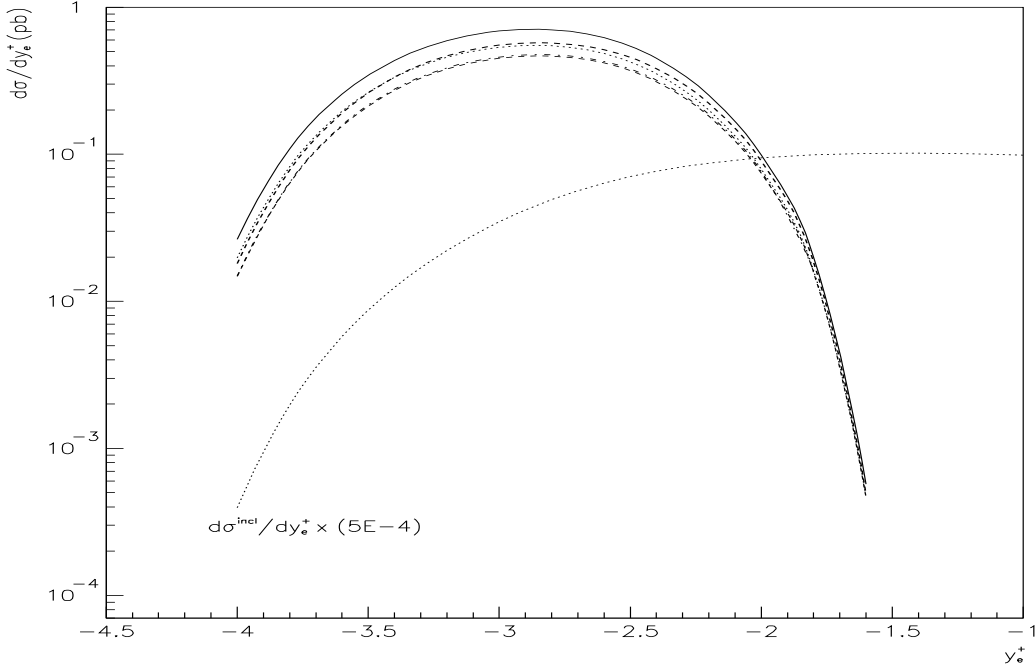,height=4in,width=6in,clip=}}
\caption{\sf Rapidity distribution of $e^{+}$ in $W^{+}$ production with the
        cut $x_{\Pomeron} < 0.01$.  The description of the various curves
    is the same as in the caption for Fig.~\ref{fig1Wm}. }
\label{fig1Wp}
\end{figure}

The corresponding cross sections for $W^{+}$ are shown in
Fig.~\ref{fig1Wp}.
The cross sections are larger than for the $W^{-}$, because a valence
up quark from the proton can be used to make a $W^{+}$, especially at
large negative rapidities.
In the plot,
the rapidity gap exists for $y_{e^{+}}>-1.6$.
The same features as in the curves of Fig.~\ref{fig1Wm} can be observed and
thus, the same general inferences for $W^-$ production can be made for
this case as well.

\subsection{Comparison to CDF data for $W$ production}

The CDF collaboration has presented data on diffractive
$W$ production from $p{\bar p}$ collisions at ${\sqrt s}=1800 \, \rm{GeV}$
\cite{CDF.wgap}.
The $W$'s are produced with a rapidity gap in the region
$2.4 < |\eta | < 4.2$.
They find that
the fraction of diffractive to
non-diffractive
$W$ production
is \cite{CDF.wgap}
$R_{W}=[1.15 \pm 0.51({\rm stat}) \pm 0.20 ({\rm syst})]\%$.
This value corresponds to diffractive data corrected up to
$x_{\Pomeron}=0.1$ \cite{KG.Moriond}.

So, in Table \ref{table:cdfWrg} we present our diffractive fractions
using Eq.~(\ref{vbincldfcs}) and our fits with $\alpha_{\Pomeron}=1.14$
for several different values of $x_{\Pomeron}^{\rm max}$.
They are computed with the diffracted hadron being allowed to be either
the proton or the antiproton.
We see that for $x_{\Pomeron}^{\rm max}=0.01$, the rates are
an order of magnitude
smaller while for $x_{\Pomeron}^{\rm max}=0.05$, the rates are
of the same order as the data.  However, the preferred fits, with
a large amount of
initial glue (B,D and SG), yield rates which are
about two to three times larger
than the data.  For $x_{\Pomeron}^{\rm max}=0.1$,
our rates are a factor
three to six
larger than the central data value.

\begin{table}
\begin{center}
    \begin{tabular}{|c!{\vrule width 2pt}c|c|c|}\hline
Fit & $x_{\Pomeron}^{\rm max} = 0.01$
 & $x_{\Pomeron}^{\rm max} = 0.05$ & $x_{\Pomeron}^{\rm max} = 0.1$ \\ \hline
      A  & 0.12\% & 1.9\% & 3.3\% \\
      B  & 0.14\% & 2.6\% & 5.1\% \\
      C  & 0.12\% & 1.8\% & 3.2\% \\
      D  & 0.18\% & 3.5\% & 6.9\% \\
      SG & 0.14\% & 2.2\% & 4.1\% \\
      \hline
    \end{tabular}
\end{center}
\caption{\sf Diffractive fractions $R_W$ for $W$ production when either
$p$ or $\bar p$ diffracts.}
\label{table:cdfWrg}
\end{table}

\section{Diffractive Jets}
\label{sec:jet.calcs}

In this section, we present our results for jet production.
We imposed the following cuts on the
jet cross sections.  These represent the effect of appropriate
experimental cuts \cite{CDF.jgap,D0.1}
and of cuts to improve the significance of the signal.
\begin{itemize}

\item
    We require that two jets are produced in the same half of the
    detector, i.e., $y_{1}y_{2}>0$, where $y_{i}$ is the rapidity of jet
    $i$.  This eliminates the region where the jets are in
    opposite hemispheres, since that region is well populated by
    non-diffractive events but is relatively unpopulated by
    diffractive events, because of the rapidity gap requirement.

\item
    Each jet is required to have a transverse energy $E_{T}$ greater
    than 20 GeV.  This ensures that we are definitely in the
    perturbative region for the jets, but the cut could be
    relaxed.

\item
    Each jet's rapidity satisfies $|y|>y_{cut}\equiv 1.8$.

\end{itemize}
Next, we integrated over the rapidity of
one of the jets
to obtain a single jet distribution, but still subject to the
above cuts on the other jet.  Equations (\ref{diffcs1})
and (\ref{nondiffcs1}) were used for the diffractive and inclusive
cross sections, respectively, with the parton distributions
evolved to the scale $E_{T}$.
For the diffractive cross sections, the $x_{{\Pomeron}}$ integral was
performed up to $x_{{\Pomeron}}^{\rm max}=0.01$.
In the following discussion, we will denote the rapidity of the final state
jet by $y_{\rm jet}$ instead of $y$.

\begin{figure}
\centerline{   \psfig{file=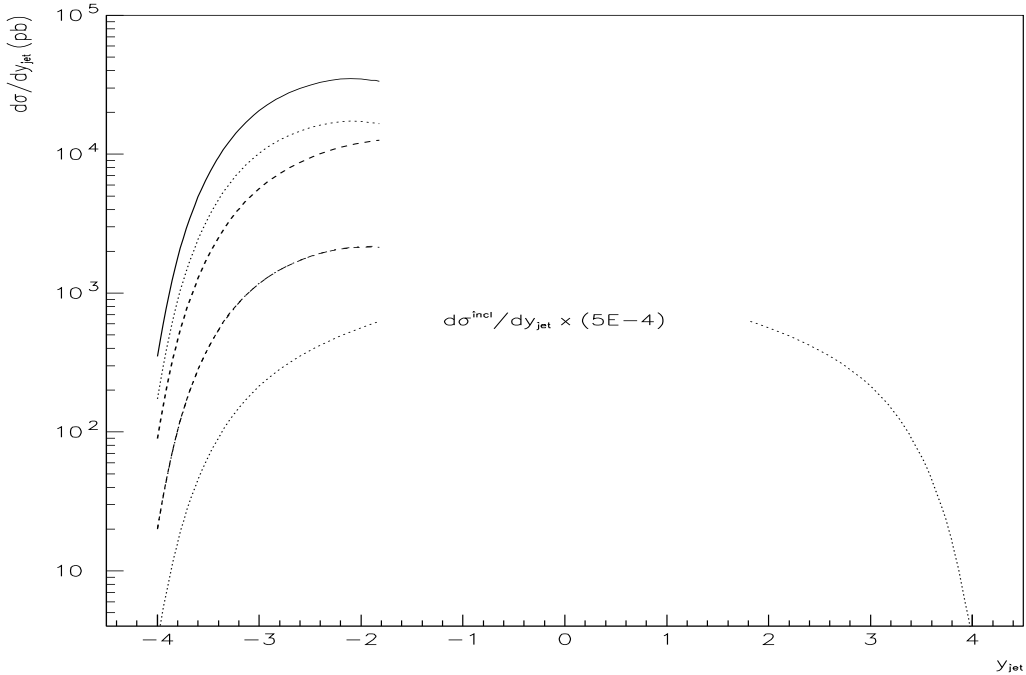,height=4in,width=6in,clip=}}
\caption{\sf Rapidity distribution of jet cross sections,
        with $E_T > 20\, {\rm GeV}$, $y>1.8$, and $x_\Pomeron <
        0.01$.
    The description of the various curves
    is the same as in the caption for Fig.~\ref{fig1Wm}. }
\label{fig.jet}
\end{figure}

The resulting cross sections are shown in Fig.~\ref{fig.jet}.  There are no
points in the
middle part of the plot because of the rapidity cut.
The cross sections using low glue fits A and C are nearly identical as
depicted by the overlapping dashed (A) and dot-dashed (C) curves in the figure.
The high glue fits D (solid curve), B (dotted) and SG (heavy dashed) yield
cross sections that are about an order of magnitude larger,
with D being largest,
than those using low glue fits.
This difference reflects the sensitivity of this
particular type of cross section to the gluon content of the Pomeron.
The lower dotted curve, which is symmetric about $y=0$,
represents the inclusive cross section
scaled down by a factor of $5 \times 10^{-4}$.

\begin{figure}
\centerline{   \psfig{file=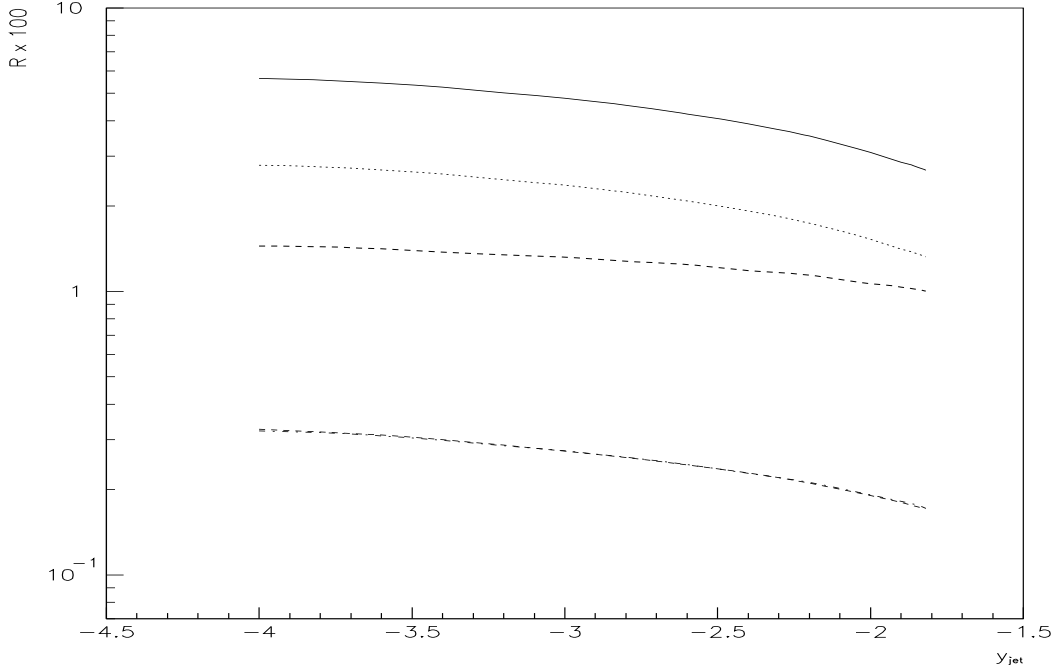,height=4in,width=6in,clip=}}
\caption{\sf Diffractive jet production percentages,
        with $E_T > 20\, {\rm GeV}$, $y>1.8$, and $x_\Pomeron <
        0.01$.
    Solid curve results from using fit D,
    upper dashed with fit SG, dotted with fit B, lower dashed with
    fit A and dot-dashed with fit C.  }

\label{fig.jetrates}
\end{figure}

The diffractive jet percentages are shown in Fig.~\ref{fig.jetrates},
where $R \times 100$ is plotted as a function of $y_{\rm jet}$, with
$R={d\sigma ^{\rm jet, diff} /dy_{\rm jet}\over
   d\sigma ^{\rm jet, incl}/dy_{\rm jet}}$.
One finds that the rates $R$ are largest when fit D is used, varying
from 2.7\% to 5.7\%.
With fit B, whose gluon distribution is about a
factor of two lower than fit D,
the rates are also about a factor of two smaller.
The rates obtained with fits A and C are much lower
ranging from 0.2\% to about 0.3\%.
With fit SG, the resulting curve is relatively flat giving a rate of
about 1.2\%.
The rates are largest at $y_{\rm jet}=-4$, then decrease as $y_{\rm jet}$
increases.
Of course, the large rates for distributions D, B and SG,
all with the large gluon distribution,
directly result from the fact that there is a gluon-induced subprocess.

We end this section by making comparisons with data
on diffractive dijet
production from CDF and D0 at ${\sqrt s}=1800 \, {\rm GeV}$.
CDF has measured dijet data both with a rapidity gap
requirement \cite{CDF.jgap} and with Roman pots \cite{CDF.Pot}
along the antiproton beam direction.
In the first case, the cross section for dijets produced opposite a
rapidity $(\eta )$ gap in the region $2.4<|\eta |<4.2$ is measured.
Each jet is required
to have a minimum $E_{T}$ of 20 GeV and rapidity
$1.8 < |\eta | < 3.5$.
They also measure the
dijet cross section without a rapidity gap, i.e., what we refer to
in this paper as the
inclusive cross section.  The diffractive fraction they
measure is \cite{CDF.jgap}
$R_{JJ}=[0.75 \pm 0.05({\rm stat}) \pm 0.09({\rm syst})]\%$.
This measured value is appropriate for $x_{\Pomeron} \leq 0.1$
\cite{KG.Moriond}.
The fractions that we obtain
using the above cuts and our fits with $\alpha_{\Pomeron}=1.14$
are shown in Table \ref{table:cdfJJrg}, for several values of
$x_\Pomeron^{\rm max}$.
Our calculation assumes that either the antiproton or the proton
is diffracted.
The rates obtained with fit D or B are from 3 to 22
times larger, while those obtained with fit C or A
range from being about 70\% smaller to being a few percent larger
than the measured value, depending on the value of
$x_{\Pomeron}^{\rm max}$.
The rates using fit SG are also significantly greater than the data but
smaller than the rates with fits B and D.
This reflects the low number of gluons in fit SG; they were more
effective in the photoproduction at producing jets because of their
relatively large fractional momentum relative to the Pomeron.

\begin{table}
\begin{center}
    \begin{tabular}{|c!{\vrule width 2pt}*{2}{D{.}{.}{6}|}{D{.}{.}{5}|}}\hline
        Fit &
        \multicolumn{1}{c|}{$x_{\Pomeron}^{\rm max} = 0.01$} &
        \multicolumn{1}{c|}{$x_{\Pomeron}^{\rm max} = 0.05$} &
        \multicolumn{1}{c|}{$x_{\Pomeron}^{\rm max} = 0.1$} \\ \hline
        A & 0.23\% & 0.63\% & 0.83\% \\
        B & 1.9 \% & 6.0 \% &  8.1 \% \\
        C & 0.23\% & 0.61\% & 0.79\% \\
        D & 3.9 \% & 12.3 \% & 16.4 \% \\
        SG & 1.2 \% & 2.5 \% & 3.2 \% \\ \hline
    \end{tabular}
\end{center}
\caption{\sf Diffractive fractions $R_{JJ} $ for dijet production
when either $p$ or $\bar p$ diffracts and using
         cuts on $E_T$ and $y$ appropriate for the CDF rapidity gap data.
}
\label{table:cdfJJrg}
\end{table}

With their Roman-pot-triggered diffractive sample, CDF
has measured a diffractive
fraction of $R_{JJ}=[0.109 \pm 0.003 \pm 0.016]\%$.
The data in this sample correspond
to $x_{\Pomeron}$ in the range $0.05<x_{\Pomeron}<0.1$, with the jets having
minimum $E_{T}$ of 10 GeV.  The fractions we obtain using the
same kinematic cuts and our fits with $\alpha_{\Pomeron}=1.14$
are presented in Table \ref{table:cdfJJpots}.
In this case, our calculation assumes that only the antiproton is diffracted.
The ones obtained with fits D, B and SG are from 8 to 34 times
larger than the data,
while those obtained with fits C and A are about
twice as large.

\begin{table}
\begin{center}
    \begin{tabular}{|c!{\vrule width 2pt}c|c|c|c|c|} \hline
        Fit     &     A &     B &     C & D & SG \\ \hline
$R_{JJ}$
       & 0.20\% & 1.8\% & 0.19\% & 3.7\% & 0.85\% \\ \hline
    \end{tabular}
\end{center}
\caption{\sf Diffractive fractions $R_{JJ}$ for dijet production
when only $\bar p$ diffracts and using
         cuts on $x_\Pomeron$, $E_T$ and $y$ appropriate to the CDF
         Roman pot data.
}
\label{table:cdfJJpots}
\end{table}

Finally, D0 also has some preliminary data \cite{D0.1} on diffractive
dijet production.  They require a
rapidity gap opposite the dijets, which
have $E_{T}^{\rm min}=12 \, {\rm GeV}$ and
$|\eta _{\rm jet}| > 1.6$.  The diffractive fraction they measure
with an estimated $x_{\mathbb P}^{\rm max}=0.03$
is $R_{JJ}=[0.67 \pm 0.05]\%$.  Our
calculated fractions are shown in Table \ref{table:d0rg}; as with our
previous calculations, we use the fits with $\alpha_{\Pomeron}=1.14$
and assume that either the antiproton or the proton is diffracted,
The realistic fits (with a large gluon content) are well above the
data, by factors of 9, 18, and 4 for fits B, D, and SG, respectively.
The cross sections obtained from fits A and C are a bit smaller than
the data; these fits give a correct normalization for diffractive DIS,
so again we see the importance of the photoproduction data in
demonstrating a breakdown of factorization.

\begin{table}
\begin{center}
    \begin{tabular}{|c!{\vrule width 2pt}c|c|c|c|c|} \hline
        Fit     &     A &     B &     C & D & SG \\ \hline
$R_{JJ}$
       & 0.59\% & 5.8\% & 0.57\% & 11.8\% & 2.4\% \\ \hline
    \end{tabular}
\end{center}
\caption{\sf Diffractive fractions $R_{JJ}$ for dijet production
when either $p$ or $\bar p$ diffracts and using
         cuts on $x_\Pomeron$, $E_T$ and $y$ appropriate for the
         D0 data.}
\label{table:d0rg}
\end{table}

\section{Conclusions}
\label{sec:concl}

We have presented parton distributions in the Pomeron resulting from
fits to data on diffractive DIS and diffractive photoproduction at
HERA.  In order to explore the sensitivity of the data to different
aspects of the parton densities, we made several fits with different
assumptions for the initial parton densities.  We find that only those
parameterizations with a large amount of glue (B, D and SG) are able
to provide a good fit to the photoproduction data.  The other two
parameterizations (A and C), which are constrained to have no gluons
in the starting distributions, badly underestimate the
photoproduction cross sections.

We also find that the normalizations of both the quark and gluon
densities in the Pomeron are well determined by the data.  As regards
the shape, hard distributions are preferred.  But in the case of the
gluon, the question still remains as to whether a conventional hard
distribution ($1-\beta$ at large $\beta$) or something harder is
correct.  We are able to obtain satisfactory fits with both a hard
gluon, in fit D, and a harder gluon, in fit SG.  We have shown how
measurements of the $\beta$ dependence of the photoproduction cross
section will be able to provide much better information.

From our fits,
we predicted the cross sections for vector
boson production and dijet production in diffractive $p{\bar p}$
interactions at the Tevatron.  The rates represent a realistic
prediction of the cross sections, {\em given the assumption of
factorization}.
We find that the predictions are a factor of several above the measured
cross sections.
In the case of the jet cross sections, it is only for
the physically correct ``high-gluon'' fits that the predictions
substantially exceed the data.
The lack of agreement between the predictions and the data indicates a
substantial breakdown of factorization in diffractive $p\bar p$
interactions.

For the predictions to match the measured diffractive rates of $W$
production by CDF, suppression factors ($\equiv$ prediction/data)
ranging from three to six must be applied.  In the case of diffractive
dijet production, the suppression factors appear to be somewhat
larger, around 10.  (We refer only to the realistic fits, with a large
amount of glue.)

Further work to measure the suppression factors is necessary to obtain
a fuller understanding of the dynamics of diffractive hadron-hadron
interactions.  One interesting possibility is to search for the
contribution predicted by the coherent Pomeron mechanism of Collins,
Frankfurt and Strikman \cite{nonfact}, which in fact gives an {\em
enhancement} of the cross section at large $\beta$.  Such an
enhancement is suggested by the UA8 data \cite{UA8}.  This and our
results on photoproduction show that the measurement of $\beta$
distributions is important.  It should be noted that the UA8 data are
at larger $|t|$ than the data which we have fitted.

With regards to extracting diffractive parton densities, further work is also
needed to understand the differences between the ZEUS
and H1 data, as illustrated in Fig.\ \ref{combo}.  The differences are
suggestive of a systematic error that is correlated point-to-point.
This indicates that we need to be careful about taking the $\chi^2$
values at face value, and in fact that systematic errors need to be
treated more correctly.
Our negative value for the soft quark term in fit D is
worrying; note that it is driven by the H1 data, as can be seen from
the $\chi^2$ values in Table \ref{Chi2.Table}.

It is also important to test the universality of $\alpha_\Pomeron$,
for example, to test whether its value is different in exclusive and
inclusive processes, as is suggested by Fig.\ \ref{allH1}.

Finally, further tests of factorization can be accomplished at HERA.  For
example, we expect hard-scattering factorization to be valid for heavy
quark production in DIS as well,
but not for any resolved photoproduction process.

{\em Note added:} After completion of the work for this paper, a paper
by the ZEUS collaboration \cite{ZEUS.fits} appeared.  It provides the
official version of the diffractive photoproduction data
\cite{ZEUS.Jerusalem} that we showed in Fig.\ \ref{photo.1994}, and
the paper reports a QCD analysis of the ZEUS data.  This analysis was
performed independently of the one in the present paper, but in a
similar style, and the conclusions as regards the parton densities in
the Pomeron are similar.

\section*{Acknowledgments}

This work was supported in part by the U.S.\ Department of Energy
under grant number DE-FG02-90ER-40577, and by the U.S. National
Science Foundation.
We are grateful for many discussions with our colleagues,
particularly those on the CTEQ and ZEUS collaborations and with
M. Albrow, A. Brandt, J. Dainton, and D. Goulianos.


\begin{references}
\bibitem{CTEQ}
    The CTEQ evolution package is part of the POMPYT-C package
    \cite{POMPYT-C}.
\end{references}
\end{document}